\PassOptionsToPackage{dvipsnames}{xcolor}
\documentclass[acmsmall,screen,authorversion,nonacm]{acmart} %

\usepackage[utf8]{inputenc}

\usepackage{csquotes}
\usepackage{float}
\usepackage{listings}
\usepackage{enumitem, hyperref}
\usepackage{mdframed}

\usepackage{caption}
\usepackage{subcaption}
\usepackage{tikz}
\usetikzlibrary{trees}
\lstdefinestyle{myCodestyle}{
  basicstyle={\ttfamily\footnotesize},
  columns=fixed,
  morestring=[b]" ,
  keywords={},
  comment=[l]{//},
  frame=none,
  aboveskip=3mm,
  belowskip=3mm,
  showstringspaces=false,
  numbers=left,
  xleftmargin=1.9em,
  numberstyle=\tiny\color{gray},
  commentstyle=\color{gray},
  breaklines=true,
  breakatwhitespace=true,
  tabsize=2,
  escapechar=§,
  backgroundcolor=\color{white},
  numberblanklines=true,
  firstnumber=last
}
\usepackage[figure,linesnumbered,noresetcount]{algorithm2e}

\usepackage{mathpartir}
\usepackage{mathtools} %

\usepackage{xspace}
\newcommand{\eauthor}{\textit{author}\xspace}
\newcommand{\eauth}{\textit{auth}\xspace}
\newcommand{\epre}{\textit{pre}\xspace}
\newcommand{\epath}{\textit{path}}
\newcommand{\eid}{\textit{id}\xspace}

\newcommand{\Revoke}[1]{\texttt{Revoke}(#1)}
\newcommand{\Update}[1]{\texttt{Update}(#1)}

\newcommand{\payload}{\mathit{op}}
\definecolor{chosencolor}{rgb}{0.65,0.48,0.0}

\newcommand{\hashof}[1]{#1^{\#}}
\newcommand{\PathDesc}{\textit{PathDesc}}
\NewDocumentCommand{\RuleName}{s O{id} m}{%
  \textnormal{\LabTirName{#3}}%
  \IfBooleanF{#1}{\ensuremath{_{\textnormal{\textit{#2}}}}}%
}
\newcommand{\Prd}{\Pi_{\mathrm{rd}}}
\newcommand{\Pwr}{\Pi_{\mathrm{wr}}}
\newcommand{\mayread}{\textit{may\_read}}
\newcommand{\maywrite}{\textit{may\_write}}
\newcommand{\mayrevoke}{\textit{may\_revoke}}
\newcommand{\causallyready}{\textit{causally\_ready}}
\newcommand{\delegationvalid}{\textit{delegation\_valid}}

\usepackage{glossaries-extra}
\usetikzlibrary{positioning, decorations.pathreplacing}
\usepackage{tikz}
\usetikzlibrary{positioning, fit, backgrounds}
\usetikzlibrary{shapes.geometric}

\definecolor{delegateColor}{RGB}{0,158,115}    %
\definecolor{revokeColor}{RGB}{213,94,0}    %
\definecolor{updateColor}{RGB}{110,110,110} %
\newcommand{\eventicon}[1]{%
  \tikz[baseline=-0.55ex]{\node[#1, draw, thick, inner sep=0pt,
                                minimum size=1.7ex] {};}}
\newcommand{\iconDelegate}{\eventicon{diamond, aspect=1, draw=delegateColor,
                                   fill=delegateColor!15}}
\newcommand{\iconUpdate}{\eventicon{circle, draw=updateColor, fill=none}}
\newcommand{\iconRevoke}{\eventicon{regular polygon, regular polygon sides=8,
                                    draw=revokeColor, fill=revokeColor!15}}

\DeclareRobustCommand{\evnamed}[2]{\mbox{\raisebox{0.25ex}{#1}\kern0.05em$#2$}}
\DeclareRobustCommand{\evDelegate}[1]{\evnamed{\iconDelegate}{#1}}
\DeclareRobustCommand{\evUpdate}[1]{\evnamed{\iconUpdate}{#1}}
\DeclareRobustCommand{\evRevoke}[1]{\evnamed{\iconRevoke}{#1}}

\usepackage{mdframed}
\newmdenv[
  topline=false, bottomline=false, rightline=false,
  linewidth=1.2pt, linecolor=black!20,
  leftmargin=0pt, rightmargin=0pt,
  innerleftmargin=8pt, innerrightmargin=0pt,
  innertopmargin=2pt, innerbottommargin=2pt,
  skipabove=0.7\baselineskip, skipbelow=0.7\baselineskip,
]{aside@frame}

\setabbreviationstyle[acronym]{long-short}
\glsdisablehyper
\newacronym{acl}{ACL}{Access Control List}
\newacronym{ardt}{ARDT}{Algebraic Replicated Data Type}
\newacronym{bec}{BEC}{Byzantine Eventual Consistency}
\newacronym{bft}{BFT}{Byzantine fault tolerant}
\newacronym{crdt}{CRDT}{Conflict-free Replicated Data Type}
\newacronym{i-confluent}{$\mathcal{I}$-confluent}{invariant confluent}

\setcopyright{none}
\begin{document}

\title{GUARDTs: Decentralized Access Control for Replicated Data Types}

\author{Christian Kuessner}
\orcid{0009-0004-0317-2649}
\affiliation{%
  \institution{Technische Universität Darmstadt}
  \country{Germany}
}
\author{Ragnar Mogk}
\orcid{0000-0003-4583-1791}
\affiliation{%
  \institution{Technische Universität Darmstadt}
  \country{Germany}
}
\author{Julian Haas}
\orcid{0000-0001-9959-5099}
\affiliation{%
  \institution{Technische Universität Darmstadt}
  \country{Germany}
}
\author{Annette Bieniusa}
\orcid{0000-0002-1654-6118}
\affiliation{%
  \institution{Rheinland-Pfälzische Technische Universität Kaiserslautern-Landau}
  \country{Germany}
}
\author{Mira Mezini}
\orcid{0000-0001-6563-7537}
\affiliation{%
  \institution{Technische Universität Darmstadt}
  \country{Germany}
}
\affiliation{%
  \institution{National Research Center for Applied Cybersecurity (ATHENE)}
  \country{Germany}
}
\affiliation{%
  \institution{Hessian Center for Artificial Intelligence (hessian.AI)}
  \country{Germany}
}

\renewcommand{\shortauthors}{Kuessner et al.}

\begin{abstract}
Replicated data types providing strong eventual consistency offer a principled
foundation for developing decentralized software.
Yet existing approaches to programming such data types leave access control unresolved: they either provide
none at all, delegate it to a trusted coordinator, reintroducing the
centralization they were designed to avoid, or tie it to specific
data-type constructions, while typically assuming non-Byzantine replicas
and static policies.

We present \textsc{GUARDTs}: \textbf{G}uarded \textbf{U}nforgeable \textbf{A}lgebraic \textbf{R}eplicated \textbf{D}ata \textbf{T}ypes,
which embed fine-grained permissions directly into the application data model,
operate without a central authority, support dynamic policy evolution,
and tolerate Byzantine replicas.
We formally prove that \textsc{GUARDTs} guarantee confidentiality,
integrity, and availability 
and show empirically that their access control adds negligible overhead on top
of the cost of Byzantine-resilient replication itself.
\end{abstract}

\maketitle

\section{Introduction}

Distributed applications increasingly rely on centralized, cloud-based 
infrastructure to synchronize state across components and to enforce 
access control. The appeal is clear: centralization hides the complexities 
of distribution and provides a single source of truth for permissions. 
Yet it concentrates control in a third party, raising concerns about data 
ownership, privacy, and security; it imposes uniform policies ill-suited 
to the diverse needs of individual applications; and it introduces a single 
point of failure that undermines availability and increases latency.

Strongly eventually consistent replicated data types, such as Conflict-free Replicated Data Types (CRDTs)~\cite{shapiroConflictFreeReplicatedData2011}, provide a principled programming foundation for decentralized 
applications.
Yet existing approaches to programming such data types leave access control unresolved. 
Widely used CRDT libraries
such as Automerge and Yjs provide
no access control at all: by default, replicas are trusted and every replica can read and write everything.
Other approaches either delegate it to a trusted authority~\citep{renauxSecureRDTsEnforcing2023}, reintroducing the
centralization they were designed to avoid, or tie it to specific
data-type constructions~\citep{raultDistributedAccessControl2022}.

To close this gap, we present an approach to access control for decentralized, strongly eventually consistent replicated data types,
called GUARDTs (\textbf{G}uarded \textbf{U}nforgeable \textbf{A}lgebraic \textbf{R}eplicated \textbf{D}ata \textbf{T}ypes).
GUARDTs guarantee the three classical goals of access control: \emph{confidentiality}, \emph{integrity}, and \emph{availability} (CIA): confidentiality ensures that no principal learns anything about data it is not permitted to read, integrity ensures that only authorized writes affect the state, and availability ensures that authorized readers eventually observe every write they are entitled to see.

In addition, our approach to access control has four properties that decentralized applications need.
First, \emph{decentralization}: the access-control mechanism operates without a central authority; a trusted coordinator that manages permissions would reinstate the very single point of trust that decentralized architectures are designed to eliminate.
Second, \emph{generality}: the mechanism neither is tied to specific data-type constructions nor treats application data as opaque; instead, it is applicable to arbitrary application data types while still being able to inspect their specific structure to enforce fine-grained policies in which read and write privileges differ across replicas and across sub-components of the data types. 
Third, \emph{policy evolution}: access rights remain updatable at runtime as users delegate permissions for tasks and revoke them upon completion, as team membership changes, or as contractors receive temporary write access.
Fourth, \emph{Byzantine resilience}: in decentralized applications, replicas may be adversarial, and the mechanism remains correct even under Byzantine behavior.

Achieving CIA and the four properties of the enforcement mechanism together is non-trivial.
Without a trusted mediator, every replica must enforce policies locally and cannot assume that its peers are honest: Byzantine replicas may impersonate others, inject unauthorized writes, or attempt to read data they are not permitted to see. At the same time, correct replicas must agree on which updates are authorized, although they observe permission changes in different orders. Revocations are particularly difficult: a revocation 
may be concurrent with writes of the replica whose rights were revoked, so replicas must consistently decide which of these writes to accept.%

We address these challenges as follows. For data generality, we build upon \glspl{ardt}~\cite{kuessnerAlgebraicReplicatedData2023}, a class of delta-based convergent algebraic data types~\cite{almeidaDeltaStateReplicated2018}. We leverage their algebraic, tree-like structure to define fine-grained permission trees, tightly integrating access control into the data model. To enable policy evolution, we model permissions as first-class capability objects that can be delegated and revoked. The authorization state is replicated as a HashDAG~\cite{kleppmannMakingCRDTsByzantine2022} among all replicas, providing each with converging, tamper-evident evidence for local authorization decisions.
Enforcement happens at the replication layer: replicas locally filter deltas based on the receiver's read permissions and materialize only deltas whose write authority can be verified. Cryptographic commitments hide delta contents from unauthorized replicas while ensuring that authorized replicas converge to identical state. Revocations 
take effect without coordination, closing access to future writes while preserving already-delivered information.

In summary, we make the following contributions.
We design GUARDTs, combining path-based capabilities, a replicated event graph, and enforcement at the replication layer, and present them informally in \autoref{sec:protocol-overview} and formally in \autoref{sec:permissions_and_enforcement}, based on the programming and system model of \autoref{sec:progSysModel}.
We prove that GUARDTs guarantee CIA despite decentralized, dynamic enforcement in a Byzantine setting (\autoref{sec:correctness}).
We show empirically that their access control adds negligible overhead on top of Byzantine-resilient replication (\autoref{chap:eval}).
Finally, we discuss related work (\autoref{chap:related work}) and future work (\autoref{chap:conclusion}).

\section{Programming and System Model}
\label{sec:progSysModel}

In this section, we present and justify our assumed application programming and system models, which motivate the design goals of our decentralized access-control framework.

\subsection{Application Programming Model with Algebraic Replicated Data Types (ARDTs)}
\label{sec:application modeling}

As a running example, consider an application for managing business trips of service engineers who inspect industrial plants at customer sites. On a trip, an engineer records the inspections of individual plant components, such as valves or pumps, and the expenses incurred along the way.
\autoref{fig:travel-adt} shows the application's data types in Scala, modeled as algebraic data types (ADTs), a common programming abstraction in modern languages.
A \texttt{TravelPlan} consists of a \texttt{title}, \texttt{inspections} of plant components, and \texttt{expenses}, a \texttt{Map} of \texttt{Expense} objects.
An \texttt{Expense} consists of a \texttt{description}, an \texttt{amount}, and an optional \texttt{comment}, represented as an \texttt{Option}.\footnote{We include the definition of \texttt{Option}, although it is a built-in Scala type, to illustrate a sum type.}

To let multiple users manage travel plans on their own devices without relying on a central server, we model \texttt{TravelPlan} and \texttt{Expense} as Algebraic Replicated Data Types (ARDTs)~\cite{kuessnerAlgebraicReplicatedData2023}, which embed conflict-free replication with strong eventual consistency into the semantics of standard algebraic data types. Akin to state-based Conflict-free Replicated Data Types (CRDTs)~\cite{shapiroConflictFreeReplicatedData2011}, ARDTs structure state as a join-semilattice with a commutative, associative, and idempotent merge.

Designing correct CRDTs by hand is notoriously error-prone, even for experts~\cite{kleppmannAssessingUnderstandabilityDistributed2022}. ARDTs overcome this challenge through compositionality: the framework provides replicated base types, such as \texttt{Text}, and derives a correct merge function for any composite ADT by recursively combining the merge functions of its fields and variants, guided by its algebraic structure. Strong eventual consistency therefore holds by construction.
Crucially, developers never write these merge functions by hand. 
When modeling application state with ARDTs, they need neither force-fit their domain models into a fixed CRDT library nor verify the correctness of custom merge functions. Instead, they express arbitrary application-specific operations over standard data types, with the type system ensuring well-formedness and the runtime guaranteeing consistency. 

\begin{figure}
\lstset{language=Scala}
\begin{lstlisting}
case class TravelPlan(
  title: Text,
  inspections: Map[UniqueId, Text],
  expenses: Map[UniqueId, Expense]
)
case class Expense(
  description: Text, amount: Text, comment: Option[Text]
)
enum Option[+T]:
  case None
  case Some(value: T)
\end{lstlisting}
\caption{Application state modeled as algebraic data types in Scala.}
\label{fig:travel-adt}

\end{figure}

ARDTs employ a variant of delta-state replication~\cite{almeidaDeltaStateReplicated2018}. Rather than transmitting the entire state upon every modification, replicas generate deltas, i.e., partial states encapsulating only the effects of recent operations. These deltas merge under the same semilattice semantics as full states in classical state-based CRDTs, enabling efficient dissemination over any transport layer guaranteeing uncorrupted message delivery.

Enforcing access control in this setting requires synchronizing deltas selectively based on replica authorization.
In our running example, Alice may want to share her inspections with her colleague Bob, who has inspected the plant before and can add notes from earlier visits. Bob may in turn want to grant the customer's shift engineer Eve access to a single inspection entry, so that she can confirm that a defect has been fixed, without involving an administrator and often without connectivity on the plant floor.
With Claudia, the office assistant who handles reimbursements, Alice wants to share only her business expenses, not her private ones. Moreover, Claudia should be permitted to append comments to expenses, but not change other fields such as amounts or descriptions. Standard CRDT replication treats a data structure as a monolithic unit, lacking awareness of such application-specific boundaries. This limitation motivates embedding access-control policies directly into the data model, so that the runtime is able to enforce differentiated read and write permissions automatically when replicating deltas.

\subsection{System and Threat Model}
\label{sec:system-model}
We consider a \emph{dynamic peer-to-peer system} in which pairs of replicas communicate asynchronously over encrypted and authenticated channels. This communication model is realistic for modern decentralized and local-first settings: replicas run on distributed devices that experience variable network latency, intermittent connectivity, and churn, yet can establish secure point-to-point channels over untrusted public networks using standard public-key cryptography (e.g., TLS or Noise-based secure sessions). 
Each replica acts as an autonomous \emph{principal} identified by a cryptographic public key. 
The system operates \emph{without a central coordinator or trusted server}. 

\emph{Replicas are mutually distrustful}. We adopt a Byzantine adversary model~\cite{lamportByzantineGeneralsProblem1982}, where faulty replicas may deviate arbitrarily from the protocol. They
may selectively drop or forge messages, introduce colluding malicious peers, or send conflicting updates to different peers to induce state divergence (equivocation). Adversaries are, however, computationally bounded: they cannot break standard cryptographic primitives, including the security of the encrypted channels, digital signatures, and collision-resistant cryptographic hash functions, which we model as random oracles.
In particular, a salted hash $H(s\,\|\,x)$ with a fresh random salt~$s$ reveals nothing about~$x$.
No replica can forge another's digital signature or find two distinct values with the same hash. We do not address scenarios where an attacker holds legitimate write permissions but acts maliciously 
(e.g., writing semantic garbage); mitigating such authorized insider threats is orthogonal to access control.

To ensure liveness, we assume \emph{eventual delivery}: non-Byzantine replicas can eventually communicate with one another, consistent with prior models for Byzantine eventual consistency~\cite{kleppmannByzantineEventualConsistency2020}. Without this assumption, an adversary could indefinitely partition the network and prevent convergence.

\section{GUARDTs in a Nutshell}
\label{sec:protocol-overview}

In this section, we provide a high-level informal overview of GUARDTs and illustrate it with the running example. Subsequently, \autoref{sec:permissions_and_enforcement} defines the protocol formally and \autoref{sec:correctness} formally proves its properties.
\subsection{Overview}

\noindent
\emph{Path-based capabilities.}
Permissions in our protocol take the form of \emph{capabilities} and are expressed using delegation events that grant a designated replica sets of \emph{path descriptors} it may read and write, following the hierarchical tree structure of ARDTs (\autoref{sec:specifying-permissions}). Every event references the capability that authorizes it, and only the delegatee of that capability may use it. A path descriptor may cover the entire data type, a specific field, or a dynamically populated collection, down to a single leaf node, giving policies the same granularity as the data model itself. Capabilities are hierarchical: granting access to a descriptor implicitly covers all its structural descendants, and write access entails read access. Capabilities thus apply to any ARDT without type-specific enforcement code.

Our protocol supports fine-grained, dynamically evolving access-control policies: the set of permissions that principals hold is not statically fixed but changes over time as replicas delegate capabilities to other replicas. Replicas can only delegate permissions they already hold, so even colluding Byzantine replicas cannot extend their authority beyond the capabilities they have been granted. Delegated capabilities can themselves be delegated further, forming \emph{delegation chains}. Moreover, replicas may revoke capabilities they have granted; revoking a capability also revokes all capabilities delegated from it, closing off entire branches of the delegation chain (\autoref{sec:revocations}). Every delegation chain originates from a designated initiating replica that initially holds full permissions over the entire path space of the replicated state. In our example, this is Alice, who owns all rights over her travel plan and delegates subsets of them to others.

\paragraph{Replicated event graph.}
Each replica must decide locally whether to apply a delta and whether to share it with a given peer.
Since delegations and revocations may be concurrent with updates and reach replicas in different orders, these decisions depend on which policy changes causally precede an update.
We therefore record all operations as events in a causally ordered, replicated event graph: \emph{delegate} events grant capabilities to a replica, \emph{update} events record application state changes, and \emph{revoke} events invalidate a delegation.
The root of the graph is a \emph{genesis event} that grants the initiating replica full read and write access to the ARDT.
Each replica uses its local view of the graph to verify delegation chains and to determine whether a revocation precedes, follows, or is concurrent with an update.
As the graph is convergent, all correct replicas eventually observe the same events and reach the same decisions.

To account for Byzantine behavior, the event graph is a HashDAG, a construction proposed for Byzantine fault-tolerant CRDTs~\cite{kleppmannMakingCRDTsByzantine2022, kleppmannByzantineEventualConsistency2020, jacobCRDTsByzantineEnvironments2022} and used by Matrix~\cite{jacobMatrixDecompositionAnalysis2020}: hash links and signatures prevent Byzantine replicas from forging events or retroactively altering them.
The graph is replicated to all replicas, but application data must remain confidential.
Update events therefore contain only a salted-hash \emph{commitment} to their delta, which is \emph{hiding} (it reveals nothing about the delta) and \emph{binding} (no replica can associate two different deltas with the same event).
The deltas themselves are kept in a separate delta store and sent only to replicas with read permission.

\paragraph{Enforcement at replication boundaries.}
Our protocol enforces policies at the replication layer, when deltas are disseminated.
For \emph{confidentiality}, a correct replica sends a delta only to peers with valid read authority for its path; for \emph{integrity}, each correct replica materializes only deltas with valid write authority.
In both cases, authority is valid if its chain of authorizing capabilities leads back to the genesis event and none of these capabilities has been revoked.
Rather than blocking writes until the existence of concurrent revocations can be ruled out, which compromises availability, writes without locally known revocations are applied optimistically and reverted if concurrent revocations become locally known.
Reads of updates committed before or concurrently with a revocation remain permitted.
This asymmetry preserves already-disclosed information while preventing replicas from circumventing revocations by backdating their writes.

\subsection{GUARDTs Illustrated}
\label{sec:materialization-example}

\autoref{fig:event-graph} shows an instance of the event graph of our running example. It contains all three kinds of events: delegations (depicted as \evDelegate{}), updates (\evUpdate{}), and revocations (\evRevoke{}); \autoref{sec:system-state} introduces them formally. The event \evDelegate{A_\bot} is the genesis event, which grants Alice access to the entire path space. Alice delegates capabilities to Bob (\evDelegate{A_2}) and to Claudia (\evDelegate{A_3}); in turn, Bob delegates capabilities to Eve (\evDelegate{B_4}). The graph in \autoref{fig:event-graph} depicts one replica's view at a single point in time; other replicas may hold different views (subsets or supersets of it). Eventually, all correct replicas receive every event shown here, possibly extended by further events.

\colorlet{delegateText}{delegateColor!60!black}
\colorlet{revokeText}{revokeColor!60!black}
\colorlet{updateText}{black}

\newcommand{\evsize}{8.5mm}
\newcommand{\evnode}[4]{%
  \node[#1] (#2) at #3 {};%
  \node[evlabel, text=#1Text] at (#2) {#4};%
}

\begin{figure}
\centering
\begin{tikzpicture}[
  x=1.5cm, y=1.25cm,
  event/.style={draw, thick, inner sep=0pt, minimum size=\evsize},
  delegate/.style={event, diamond, aspect=1, draw=delegateColor, fill=delegateColor!15},
  revoke/.style={event, regular polygon, regular polygon sides=8,
                 draw=revokeColor, fill=revokeColor!15},
  update/.style={event, circle, draw=updateColor, fill=none},
  evlabel/.style={font=\footnotesize, inner sep=0pt},
  pre/.style={-latex, thick, shorten >=1pt, shorten <=1pt},
  auth/.style={-latex, densely dotted, shorten >=1pt, shorten <=1pt, gray},
  revedge/.style={-, thick, dashed, revokeColor,
                  shorten >=6pt, shorten <=1pt},
  revtip/.style={pos=1, inner sep=0pt, text=revokeColor, font=\small}
  ]

  \evnode{delegate} {a0}{(0,0)}{$A_\bot$}
  \evnode{update}{a1}{(1,0)}{$A_1$}
  \evnode{delegate} {a2}{(2,0)}{$A_2$}
  \evnode{delegate} {a3}{(3,0)}{$A_3$}
  \evnode{revoke}{a8}{(6,0)}{$A_8$}
  \evnode{delegate} {b4} {(4,1)}{$B_4$}
  \evnode{update}{c5}{(4,-1)}{$C_5$}
  \evnode{update}{e6}{(5,2)}{$E_6$}
  \evnode{update}{e7}{(6,2)}{$E_7$}
  \evnode{update}{e9}{(7,2)}{$E_9$}

  \draw[pre] (a1)  -- (a0);
  \draw[pre] (a2)  -- (a1);
  \draw[pre] (a3)  -- (a2);
  \draw[pre] (b4)  -- (a3);
  \draw[pre] (c5)  -- (a3);
  \draw[pre] (e6)  -- (b4);
  \draw[pre] (e7)  -- (e6);
  \draw[pre] (a8)  -- (c5);
  \draw[pre] (a8)  -- (e6);
  \draw[pre] (e9) -- (a8);
  \draw[pre] (e9) -- (e7);

  \draw[auth] (a1)  to[bend left=35]  (a0);
  \draw[auth] (a2)  to[bend left=40]  (a0);
  \draw[auth] (a3)  to[bend left=45]  (a0);
  \draw[auth] (b4)  to[bend right=30] (a2);
  \draw[auth] (e6)  to[bend left=25]  (b4);
  \draw[auth] (e7)  to[bend left=30]  (b4);
  \draw[auth] (c5)  to[bend left=30]  (a3);
  \draw[auth] (a8)  to[bend right=15] (a2);

  \draw[revedge] (a8) to[bend left=25] node[revtip] {$\times$} (b4);
  \draw[auth] (e9) to[bend left=45] (b4);

  \begin{scope}[shift={(0,0.95)}, font=\scriptsize]
    \node[delegate, scale=0.62]  (lg) at (0,1.15) {};
    \node[right=1.5mm of lg] {delegate};
    \node[revoke, scale=0.62] (lr) at (0,0.6) {};
    \node[right=1.5mm of lr] {revoke};
    \node[update, scale=0.62] (lu) at (0,0.05) {};
    \node[right=1.5mm of lu] {update};
    \draw[pre]  (1.9,1.15) -- (1.35,1.15);
    \node[anchor=west] at (2,1.15) {\epre};
    \draw[auth] (1.9,0.6) -- (1.35,0.6);
    \node[anchor=west] at (2,0.6) {\eauth};
    \draw[revedge] (1.9,0.05) -- node[revtip] {$\times$} (1.35,0.05);
    \node[anchor=west] at (2,0.05) {revokes};

  \end{scope}

\end{tikzpicture}
\caption{%
  An example event graph, with one row per author. Events are drawn as
  diamonds (delegate), octagons (revoke), and circles (update), and are named
  after the initial of their author -- Alice (A), Bob (B), Claudia (C),
  Eve (E).
  Solid arrows are \epre pointers referencing an event's causal predecessors;
  dotted arrows are \eauth pointers referencing the capability that authorizes
  an event; the dashed edge marks a revoked capability.
  \autoref{sec:materialization-example} walks through the example.
}
\label{fig:event-graph}

\bigskip
\centering
\begin{tikzpicture}[
  event/.style={draw, thick, inner sep=0pt, minimum size=8.5mm},
  delegate/.style={event, diamond, aspect=1, draw=delegateColor, fill=delegateColor!15},
  update/.style={event, circle, draw=updateColor, fill=none},
  evlabel/.style={font=\footnotesize, inner sep=0pt},
  auth/.style={-latex, densely dotted, thick, shorten >=2pt, shorten <=2pt, gray},
  perm/.style={font=\scriptsize, align=center, inner sep=0pt, text width=3.1cm},
  ]

  \foreach \name/\kind/\x/\lab in {%
    a0/delegate/0/{$A_\bot$}, a2/delegate/3.5/{$A_2$},
    b4/delegate/7/{$B_4$}, e6/update/10.5/{$E_6$}}{%
    \node[\kind] (\name) at (\x,0) {};
    \node[evlabel] at (\name) {\lab};
  }

  \foreach \from/\to in {a2/a0, b4/a2, e6/b4}{%
    \draw[auth] (\from) -- node[above, font=\scriptsize, text=gray] {\eauth} (\to);
  }

  \node[perm, below=2mm of a0] {author: Alice\\delegatee: Alice\\
    $\Prd = \Pwr = \{\texttt{*}\}$};
  \node[perm, below=2mm of a2] {author: Alice\\delegatee: Bob\\
    $\Prd = \Pwr =
     \{\texttt{\textcolor{RoyalBlue}{inspections}}\}$};
  \node[perm, below=2mm of b4] {author: Bob\\delegatee: Eve\\
    $\Prd = \Pwr =
     \{\texttt{\textcolor{RoyalBlue}{inspections}.valve3}\}$};
  \node[perm, below=2mm of e6] {author: Eve\\
    writes\\
    \texttt{\textcolor{RoyalBlue}{inspections}.valve3}};
\end{tikzpicture}
\caption{The delegation chain authorizing Eve's update \evUpdate{E_6} of
  \autoref{fig:event-graph}. Each delegation is authorized by the one to its
  left and delegates a subset of its permissions, down to the single
  inspection entry that \evUpdate{E_6} writes.}
\label{fig:delegation chain}
\end{figure}

We walk through the graph of \autoref{fig:event-graph} and determine for each update whether it is authorized and thus included in the materialized state (\autoref{sec:system-state} and \autoref{sec:read-write-semantics} formalize these decisions).
The simplest case in \autoref{fig:event-graph} is the update \evUpdate{A_1}, which is authorized by the genesis event \evDelegate{A_\bot}. Claudia's update \evUpdate{C_5} is authorized by \evDelegate{A_3} and is concurrent with Bob's delegation \evDelegate{B_4} to Eve. The update event \evUpdate{E_6} has a longer delegation chain, which is depicted in \autoref{fig:delegation chain} together with the information captured in delegation events, such as the author of the delegation, the delegatee, and the permissions each delegation passes on.

The delegation chain in \autoref{fig:delegation chain} starts at the genesis event \evDelegate{A_\bot}, which captures that Alice is both the author and the delegatee and that she is granted read and write access to the entire travel plan ($\{\texttt{*}\}$).
She asks Bob to help with the inspections and authors \evDelegate{A_2}, a delegation event with $\textit{delegatee}\colon \text{Bob}$, $\Prd = \Pwr = \{\texttt{inspections}\}$.
This delegation is valid: both sets are covered by $\{\texttt{*}\}$, so Bob receives no more than Alice holds.
Bob can therefore read and modify every inspection entry, but nothing else, in particular none of the expenses.
Bob, in turn, asks Eve, the customer's shift engineer, to confirm that a defective valve has been fixed.
Because Bob is the designated delegatee of \evDelegate{A_2}, he is authorized to issue a delegation that references \evDelegate{A_2} as its authorizing parent.
He authors \evDelegate{B_4} with $\textit{delegatee}\colon \text{Eve}$ and $\Prd = \Pwr = \{\texttt{inspections.valve3}\}$, the single entry Eve is supposed to confirm. This delegation is valid, as $\texttt{inspections.valve3} \sqsubseteq \texttt{inspections}$.
Eve thus receives access to one inspection entry, with no authority over the rest of the travel plan, and her update \evUpdate{E_6} writes to exactly that path. The chain of \evUpdate{E_6} is valid.

When Alice learns that Eve's tablet has been stolen from the plant floor, she revokes Bob's delegation to Eve, \evDelegate{B_4}, with \evRevoke{A_8}, which is authorized by \evDelegate{A_2}.
Alice has already seen Eve's first update, but not her second: \evRevoke{A_8} references \evUpdate{E_6} and \evUpdate{C_5} as predecessors, but not \evUpdate{E_7}.
Eve's second update is thus concurrent with the revocation, \evUpdate{E_7} $\parallel$ \evRevoke{A_8}, which makes \evUpdate{E_7} unauthorized.
\evUpdate{E_6} is authorized by the same delegation event \evDelegate{B_4} that was revoked by \evRevoke{A_8}, but \evUpdate{E_6} $\prec$ \evRevoke{A_8}, so Alice has acknowledged \evUpdate{E_6} in her revocation.
A revocation does not affect earlier events, and \evUpdate{E_6} therefore remains authorized.
The materialized state of \autoref{fig:event-graph} thus contains \evUpdate{A_1}, \evUpdate{C_5}, and \evUpdate{E_6}, but not \evUpdate{E_7} and \evUpdate{E_9}.
\evUpdate{E_9} is a special case: it references \evRevoke{A_8} as a predecessor but is still authorized by the revoked \evDelegate{B_4}. Its author knew about the revocation, so \evUpdate{E_9} is a Byzantine event that no correct replica would create, and none will ever include the update it captures.

\section{Formalization of GUARDTs}
\label{sec:permissions_and_enforcement}
\label{sec:ardt subtrees as resources}

In this section, we formally define GUARDTs.
Capabilities grant permissions over paths of the ARDT (\autoref{sec:specifying-permissions}); delegations, revocations, and updates form a replicated event graph (\autoref{sec:system-state}); and from this graph, each replica locally decides which deltas it may send and apply (\autoref{sec:read-write-semantics}).

\subsection{Path-Based Capabilities}
\label{sec:specifying-permissions}

\begin{figure}
\centering
\begin{minipage}[c]{0.38\columnwidth}
\lstset{language=Scala}
\begin{lstlisting}
case class Expense(
  §\textcolor{RawSienna}{description}§: Text,
  §\textcolor{RoyalBlue}{amount}§: Text,
  §\textcolor{Orange}{comment}§: Option[Text]
)
enum Option[+T]:
  case §\textcolor{RoyalBlue}{None}§
  case §\textcolor{RawSienna}{Some}§(§\textcolor{Orange}{value}§: T)
\end{lstlisting}
\end{minipage}\hfill
\begin{minipage}[c]{0.6\columnwidth}
\centering
\begin{tikzpicture}[
  grow=down,
  sibling distance=2.4cm,
  level distance=1.2cm,
  edge from parent/.style={draw, -latex},
  every node/.style={align=center, font=\ttfamily\footnotesize, text=gray}
  ]

  \node {Expense}
    child { node {Text} edge from parent node[left, text=RawSienna] {description} }
    child { node {Text} edge from parent node[left, text=RoyalBlue] {amount} }
    child { node {Option}
      child { node {Some}
          child { node {Text} edge from parent node[left, text=Orange] {value} }
          edge from parent node[left, text=RawSienna] {Some}
      }
      child { node {None} edge from parent node[right, text=RoyalBlue] {None} }
      edge from parent node[right, text=Orange] {comment}
    };

\end{tikzpicture}
\end{minipage}
\caption{Type definitions of \texttt{Expense} and \texttt{Option} (left, from \autoref{fig:travel-adt}) and the resulting tree representation (right): field names and variant tags become edge labels, i.e., path segments.}
\label{fig:Expense Tree}
\end{figure}
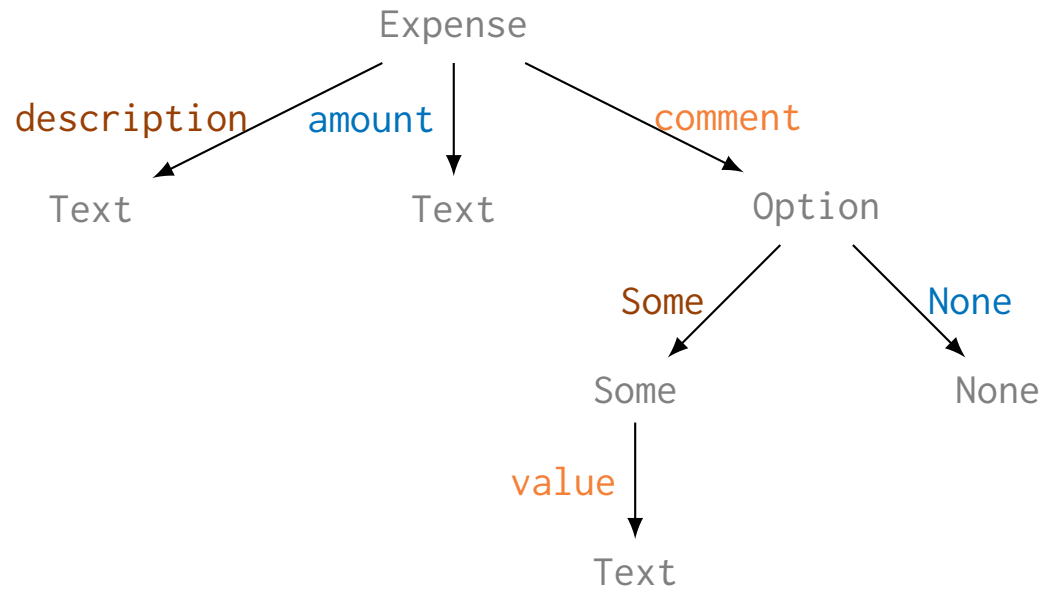

Because ARDTs are constructed via nested algebraic data types (products, sums, and collections), their structural schema forms a hierarchical tree: edges carry the field selectors of product types, and sum types branch into their variants.
\autoref{fig:Expense Tree} shows this tree for \texttt{Expense}.
Nodes of an ARDT tree can be addressed by \emph{path descriptors} and \emph{paths}. 

\begin{definition}[Path Descriptors and Paths]
\label{def:path-descriptor}

A \emph{path descriptor} is a dot-separated list of \emph{path segments} $s$, which are either the wildcard \texttt{*} or an edge label of the ARDT tree, i.e., a field name of a product type, a variant tag of a sum type, or a key of an associative collection:
\[
  \PathDesc ::= s\,(\texttt{.}\,s)^{*}
\]
A descriptor with no wildcard is a concrete \emph{path} $p$, addressing exactly one node of the ARDT tree by the sequence of edge labels traversing the tree from the root to that node. The empty path $\epsilon$ denotes the root of the ARDT and is only matched by the path descriptor \texttt{*}.
\end{definition}

Path descriptors are the basis of permissions. A \emph{permission} is the right to read or to write the nodes matched by a path descriptor. A \emph{capability} is a delegation event (\autoref{sec:system-state}) that grants its delegatee read permissions for a set of path descriptors $\Prd \subseteq \PathDesc$ and write permissions for a set $\Pwr \subseteq \PathDesc$. Every event references the delegation event that authorizes it.

A path $p$ matches a descriptor $pd$, written $p \sqsubseteq pd$, if $pd$ is a prefix of $p$, where a wildcard segment matches any single path segment.
Concrete paths thus grant permissions as fine-grained as the data type itself, while wildcards cover dynamically populated collections, such as maps with generated keys, whose elements cannot be enumerated in advance.
Because path descriptors are defined over the structural schema of any ARDT rather than a fixed set of types, the same mechanism applies to any ARDT without type-specific enforcement code.

For example, in the travel plan of \autoref{fig:travel-adt}, the path \path{expenses.e17.description} addresses the description of the expense with the (randomly generated) ID \texttt{e17}, and \path{expenses.e17.comment.Some.value} the text of its comment, if present.
The descriptor \texttt{expenses.*.description} instead matches the description of \emph{every} expense: the wildcard stands for an arbitrary key of the \texttt{expenses} map, so the descriptor also covers expenses added later.
A replica can thus delegate access to \texttt{expenses.*.comment} without enumerating expense IDs.

Each delta value $\delta$ modifies the node at a single path, written $\epath(\delta)$.
Since deltas and permissions both refer to nodes by path, deciding whether a capability covers a delta reduces to path matching.
Because matching is by prefix, access to a descriptor extends to all its structural descendants; and because modifying a data element requires observing its context, write access implies read access.
A replica is thus authorized to
\[
  \text{write } \delta \;\text{ if }\; \exists pd \in \Pwr : \epath(\delta) \sqsubseteq pd,
  \qquad
  \text{read } \delta \;\text{ if }\; \exists pd \in \Prd \cup \Pwr : \epath(\delta) \sqsubseteq pd.
\]

\subsection{Tracking and Evolving System Knowledge}
\label{sec:system-state}

The replicated event graph records all delegations, revocations, and
application state updates as causally ordered events and is implemented as a HashDAG.
In this subsection, we introduce the structure of the events, the event graph, and the actions that 
change and replicate its state.

\subsubsection{Events, Event Graph, and System Knowledge}

\begin{definition}[Replica Identifier]
\label{def:replica-id}
Let $\mathbb{I}$ be the universe of \emph{replica identifiers}.
An identifier $\mathit{id} \in \mathbb{I}$ is a cryptographic public key whose
corresponding private key is held exclusively by that replica.
\end{definition}

A replica signs every event it authors with its private key, which allows any
other replica to authenticate events without a trusted third party.
A replica identifier may correspond to an individual user, a specific device owned
by a user (allowing different devices of the same user to hold disjoint privileges),
or an isolated process that follows the principle of least privilege, i.e., holds only the permissions it needs.
Because our access-control model is agnostic to the operational boundary of a
principal, treating all authorization endpoints uniformly as replica identifiers simplifies the
formal semantics without loss of generality.

\begin{definition}[Cryptographic Hash]
\label{def:hash}
Events refer to each other, and to the data they control, only by cryptographic
hash. We write $\hashof{e}$ for the hash of a value~$e$ and, for a set~$E$,
$\hashof{E} = \{\hashof{e} \mid e \in E\}$ for the set of hashes of its
elements.
Because we use hashes as cryptographically verified pointers, we generally assume that we can use a hashed value $\hashof{e}$ as if it were the value~$e$ itself, except where noted otherwise.
\end{definition}

\begin{definition}[Event]
\label{def:event}
Let $\mathbb{E}$ be the domain of system events.
An event $e \in \mathbb{E}$ is a tuple
\[
  e \;=\; (\, \payload,\; \mathit{author},\; \mathit{auth},\; \mathit{pre} \,)
\]
of an operation $\payload \in \mathit{Operation}$ that the event performs, the
replica $\mathit{author} \in \mathbb{I}$ that authored $e$ and whose signature
authenticates it, the hash
$\mathit{auth} \in \hashof{\mathbb{E}}$ of the delegation event
\emph{authorizing} $e$, and the set
$\mathit{pre} \subseteq \hashof{\mathbb{E}}$ of hashes of the \emph{direct
causal predecessors} of $e$.
Because $\mathit{auth}$ and $\mathit{pre}$ are hashes rather than the events
themselves, altering an event invalidates the references of every event that
follows it: a Byzantine replica cannot splice, reorder, or retroactively insert
entries without detection.
This is what gives the event graph Byzantine Eventual Consistency
(BEC)~\cite{kleppmannByzantineEventualConsistency2020}.
\end{definition}

\begin{definition}[Operation]
\label{def:operation}
An operation $\payload \in \mathit{Operation}$ takes one of three forms:
\begin{itemize}
  \item[\iconDelegate] $\mathsf{Delegate}(\textit{delegatee}, \Prd, \Pwr)$
        delegates to replica $\textit{delegatee} \in \mathbb{I}$ the
        read permissions $\Prd \subseteq \PathDesc$ and
        write permissions $\Pwr \subseteq \PathDesc$.
        We call events carrying such an operation
	\emph{delegation events} and use them as capabilities for authorization.
  \item[\iconUpdate] $\mathsf{Update}(\hashof{(\eid\,\|\,\delta)})$
        establishes a cryptographic commitment (\autoref{sec:commitments})
        by \eid to ARDT delta value~$\delta$.
        We call events carrying such an operation \emph{update events}.
  \item[\iconRevoke] $\mathsf{Revoke}(\hashof{e_d})$
        revokes the delegation event~$e_d$, which it identifies by its hash.
        We call events carrying such an operation \emph{revocation events}.
\end{itemize}
\end{definition}

The domain $\mathbb{E}$ additionally contains one distinguished
\emph{genesis event}~$e_\bot$ (Definition~\ref{def:genesis}).
This event bootstraps full authority for the initiating replica~$\eid_\bot$, granting it
read and write access to the
entire path space ($\Prd = \Pwr = \{\texttt{*}\}$). %
It is axiomatically authorized ($\eauth = \hashof{0}$) and has no causal predecessors ($\epre = \emptyset$).
As the root of all other events, the genesis event is initially known to all replicas. The initiating replica thus acts as an omnipotent administrator: all authority in the system derives from it, and, as the genesis event cannot be revoked, its authority is permanent.

\begin{definition}[Genesis Event]
\label{def:genesis}
Every event graph is seeded with the same \emph{genesis event}~$e_\bot$:
\[
  e_\bot
  \;=\;
  (\,
    \mathsf{Delegate}(id_\bot,\; \{\texttt{*}\},\; \{\texttt{*}\}),\;
    id_\bot,\;
    \hashof{0},\;
    \emptyset
  \,)
\]
\end{definition}

\begin{definition}[Event Graph]
\label{def:event-graph}
An \emph{event graph} $E \subseteq \mathbb{E}$ is the set of events a replica has
created locally or received from peers.
Its edges are the predecessor references: each event $e' \in E$ has an edge to
every event $e \in E$ with $\hashof{e} \in e'.\mathit{pre}$.
We write $e \prec e'$ and say that $e$ \emph{happened before} $e'$ if $e$ is
reachable from $e'$ along these edges, and we write $e \parallel e'$ and call
the two events \emph{concurrent} if neither $e \prec e'$ nor $e' \prec e$ holds.
Reachability is a partial order on~$E$.
\end{definition}

The happened-before order may be incomplete: a Byzantine author can omit known events from the predecessors of its own events and thereby \emph{backdate} them, making them appear concurrent with events they actually follow. Conversely, hash links prevent the order from containing false precedences: an event can only reference events that already exist, so $e \prec e'$ implies that $e$ was indeed created before~$e'$.

Every ARDT update is encapsulated in a delta value that modifies a specific node in its tree. 
Replicas store the deltas they have produced and received in their local delta store. 

\begin{definition}[Delta Value and Delta Store]
\label{def:delta-store}
Let $\mathbb{D}$ be the join-semilattice of an ARDT, with merge operator~$\bigsqcup$.
A \emph{delta value} is an element $\delta \in \mathbb{D}$ that targets a single concrete path, written $\epath(\delta)$.
A \emph{delta store} $\Delta \subseteq \mathbb{D}$ is the set of concrete delta
values a replica has produced locally or received from remote peers.
\end{definition}

\begin{definition}[Local and Global Knowledge]
\label{def:knowledge}
The \emph{local knowledge} of a replica~$\mathit{id}$ is the triple
$\langle E,\, \Delta,\, I \rangle$ of its event graph, its delta store, and its
\emph{inbox} $I \subseteq \mathbb{E} \cup \mathbb{D}$ of messages (events or deltas) that
have been sent to it but not necessarily received yet.
The \emph{global system knowledge} is a finite map
\[
  K \;:\; \mathbb{I} \rightharpoonup
          \bigl(\mathit{events} : \mathcal{P}(\mathbb{E}),\; \mathit{deltas} : \mathcal{P}(\mathbb{D}),\; \mathit{inbox} : \mathcal{P}(\mathbb{E} \cup \mathbb{D})\bigr)
\]
with $K[\mathit{id} \mapsto \langle E,\, \Delta,\, I \rangle]$ denoting that
replica~$\mathit{id}$ holds event graph~$E$, delta
store~$\Delta$, and inbox~$I$.
\end{definition}

\subsubsection{Transition Rules of the Event Graph}
\label{sec:action-rule}
The event graph of a replica is modified by the rules $\RuleName*{Action}$, $\RuleName*{Send Event}$, and $\RuleName*{Receive Event}$ below.
Each rule is a step $\xrightarrow{}$ taken by a single acting replica~$\eid$. $\RuleName*{Action}$ and $\RuleName*{Receive Event}$ change only the local knowledge $K(\eid)$, while \RuleName*{Send Event} adds to the inbox of a peer $\eid_r$.

\vspace*{3ex}
\begin{flushleft}
$\inferrule*[left={\makebox[6.5em][l]{Action}}]{
  e = (\payload, \eid, \hashof{e_d}, \mathsf{heads}(E))
  \\ e_d \in E
  \\ \Delta' = \Delta \cup \{ \delta \mid \payload = \Update{\hashof{(\eid\,\|\,\delta)}} \}
}{
  K[\eid \mapsto \langle E,\, \Delta,\, I \rangle]
  \;\xrightarrow{\eid}\;
  K[\eid \mapsto \langle E \cup \{e\},\, \Delta',\, I \rangle]
}$
\\[3ex]
$\inferrule*[left={\makebox[6.5em][l]{Send Event}}]{
  e \in E
}{
  K[\eid \mapsto \langle E,\, \Delta,\, I \rangle,\;
    \eid_r \mapsto \langle E_r,\, \Delta_r,\, I_r \rangle]
  \;\xrightarrow{\eid}\;
  K[\eid_r \mapsto \langle E_r,\, \Delta_r,\, I_r \cup \{e\} \rangle]
}$
\\[3ex]
$\inferrule*[left={\makebox[6.5em][l]{Receive Event}}]{
  e \in I
  \\ \causallyready(e, E)
}{
  K[\eid \mapsto \langle E,\, \Delta,\, I \rangle]
  \;\xrightarrow{\eid}\;
  K[\eid \mapsto \langle E \cup \{e\},\, \Delta,\, I \rangle]
}$
\end{flushleft}

\vspace{5mm}
\paragraph{$\RuleName*{Action}$} This rule models a replica taking a local action that modifies its local copy of the event graph by appending a new event to it. The operation is freely chosen by the replica: honest replicas construct events that adhere to the correctness properties we discuss later in this section (Sections~\ref{sec:delegations} and~\ref{sec:read-write-semantics}), but Byzantine replicas may produce arbitrary operations.
The \eauthor of the event is the acting replica~$\eid$, which cryptographically signs the event, so that Byzantine replicas cannot forge it.
Its \eauth is the capability that the acting replica picks to authorize the event, i.e., a delegation event~$e_d \in E$ that grants the required permissions.
Its \epre is the set of heads of the local event graph~$E$, i.e., the events not in \epre of events in $E$.
Since $e_d \in E$, this implies $e_d \prec e$.
While hash-linking prevents claiming unseen predecessors, a Byzantine replica may omit known heads to backdate the event and thus subvert the protocol.
When the operation is an $\mathsf{Update}$, the delta $\delta$ it references is added to the delta store.

\paragraph{$\RuleName*{Send Event}$ and $\RuleName*{Receive Event}$} These rules model peer-to-peer dissemination of the event graph.
In $\RuleName*{Send Event}$, the acting replica~$\eid$ puts an event~$e$ from its event graph into the inbox~$I_r$ of some peer~$\eid_r$.
In $\RuleName*{Receive Event}$, the acting replica~$\eid$ takes an event~$e$ from its inbox~$I$ and inserts it into its event graph.
\RuleName*{Send Event} has no access-control guard: events contain only causal links, delegations, and hiding commitments ($e.\payload = \Update{\hashof{(e.\eauthor\,\|\,\delta)}}$), so they reveal no confidential application data and are replicated without access-control filtering.
Since the sender may be Byzantine, a received $e$ may be any event, but signatures and hash links ensure that it cannot impersonate other replicas or alter existing events.
Metadata needs to be replicated in a causally consistent manner, 
so that the events a received event points to are already present at the receiving replica.
Any algorithm for causal broadcast achieves this, and we abstract over the choice with the predicate $\causallyready(e, E)$, checked by the receiver.
We only require that $\causallyready(e, E)$ implies that all predecessors $\epre$ of $e$ are in $E$, and that $\epre \neq \emptyset$ and $e.\eauth \prec e$ unless $e = e_\bot$.
The former is necessary because an event without predecessors would otherwise constitute a second root of the event graph, which a Byzantine replica could use to create events unrelated to the genesis event.
The latter ensures that the authorizing capability of an event, and transitively its whole delegation chain, is part of the causal past of the event, and thus known to every replica that knows the event.

\subsubsection{Validating and Invalidating Delegations}
\label{sec:delegations}
\label{sec:revocations}

\begin{figure}[b]
\small
\begin{gather*}
\begin{aligned}
\delegationvalid(e_d) :=\;
           & e_d = e_\bot \;
  \vee\; \bigl( \delegationvalid(e_d.\eauth) && \text{(i)} \\
           & \qquad \wedge\; e_d.\eauthor = e_d.\eauth.\payload.\textit{delegatee} && \text{(ii)} \\
           & \qquad \wedge\; e_d.\payload.\Prd \sqsubseteq e_d.\eauth.\payload.\Prd \;
           \wedge\; e_d.\payload.\Pwr \sqsubseteq e_d.\eauth.\payload.\Pwr \bigr) && \text{(iii)} \\
\textit{revocations}(E, e_\bot) :=\; & \emptyset \\
\textit{revocations}(E, e_d) :=\;
  & \{e_r \in E \mid e_r.\payload = \Revoke{\hashof{e_d}}
        \;\wedge\; \mayrevoke(e_r)\}
    \;\cup\; \textit{revocations}(E, e_d.\eauth)\\
\eauth^*(e_\bot) :=\; & \{e_\bot\} \\
\eauth^*(e_d) :=\; & \{e_d\} \cup \eauth^*(e_d.\eauth) \\
\mayrevoke(e_r)
  :=\; & e_r.\eauthor = e_r.\eauth.\eauthor
     \;\wedge\; e_r.\eauth \in \eauth^*(e_r.\payload)
     \;\wedge\; e_r.\payload \neq \Revoke{\hashof{e_\bot}}
\end{aligned}
\end{gather*}
\caption{The delegation and revocation predicates: $\delegationvalid$, $\textit{revocations}$, and $\mayrevoke$. The authorization chain $\eauth^*(e_d)$ contains $e_d$ and all capabilities reachable from it by following \eauth pointers.}
\label{fig:delegation-validity-revocation-predicates}
\end{figure}

The predicates in \autoref{fig:delegation-validity-revocation-predicates} formalize the validity of a delegation
($\delegationvalid(e_d)$) and the effect of invalidating, i.e., revoking, an existing delegation ($\textit{revocations}$).

\paragraph{Validating Delegations.}
The $\delegationvalid(e_d)$ predicate in \autoref{fig:delegation-validity-revocation-predicates}
formalizes a valid delegation chain from a
delegation event~$e_d$ back to the genesis event~$e_\bot$. The genesis event is valid by axiom; any other delegation is validated recursively along its chain:
(i)~The authorizing parent $e_d.\eauth$ must itself have a valid delegation chain; this recursion terminates at the genesis event.
(ii)~The author of the new delegation $e_d$ must be the replica that the parent delegates access to, its \textit{delegatee}.
(iii)~The delegated read and write paths must be covered ($\sqsubseteq$) by those of the parent.
Together, (ii) and (iii) ensure that a replica can only pass on permissions it holds itself, and never more.

\paragraph{Cascading Revocations.}
The right to revoke a delegation strictly remains with the delegator: one can only revoke access one has delegated oneself.
This includes delegations that transitively depend on one's own: a replica may revoke any delegation whose authorization chain passes through a delegation it authored ($\mayrevoke$).
The genesis event cannot be revoked, as it is the root of all authority. Invalidating a delegation~$e_d$ also invalidates every delegation whose authorization chain passes through~$e_d$.
This closes the security gap that would otherwise emerge if a revoked principal could continue exercising authority via downstream sub-delegations. The $\textit{revocations}$ function in \autoref{fig:delegation-validity-revocation-predicates} computes the set of all revocations for a given delegation event $e_d$. It is recursive: the revocations for $e_d$ also include the revocations for its authorizing parent $e_d.\eauth$. It terminates because the event graph is acyclic, so the recursion eventually reaches the genesis event~$e_\bot$.

\subsubsection{Cryptographic Commitments}
\label{sec:commitments}
Every replica performs authorization checks based on the shared event graph, which requires that it
is replicated to all replicas, including those not permitted to read the data it records.
Delta values are therefore kept separately and sent only to authorized replicas, while each update event only contains a \emph{commitment} to its delta.
The commitment must be \emph{hiding} (confidentiality): it reveals nothing about the delta, so a replica that lacks read permissions can replicate the update event~$e_u$ without learning the underlying state mutation.
It must also be \emph{binding} (integrity and non-equivocation):
no replica, not even the author, can find a second delta $\delta' \neq \delta$ matching the commitment in~$e_u$, so every authorized replica that receives~$\delta$ can verify it against the event graph, and correct replicas converge on identical materialized state.

We achieve both properties with a salted hash.
A plain hash of the delta would not suffice, as it would let an unauthorized replica guess the actual payload and then confirm the correctness of the guess by checking the hash.
Update events therefore record a \emph{salted} hash as a commitment:
$e_u.\payload = \mathsf{Update}\left(\hashof{(s \,\|\, e_u.\eauthor \,\|\, \delta)}\right)$, where the salt~$s$ is a random
secret that the author chooses when creating the update and sends along with~$\delta$.
We assume that correct replicas choose fresh salts, i.e., salts never collide for any two commitments from correct replicas.
The commitment is also bound to the author, so that no other replica can replay the public commitment in an event of its own.
Under the cryptographic assumptions of our threat model (\autoref{sec:system-model}), this construction is hiding, since hashes are modeled as random oracles and $s$ is fresh and random, and binding, since hashes are collision-resistant.
As the salt is irrelevant for the remainder of the paper, we treat it as part of~$\delta$ and write $\hashof{(e_u.\eauthor \,\|\, \delta)}$.

\subsection{Enforcing Read and Write Access}
\label{sec:read-write-semantics}

In a centralized access-control system, all events are totally ordered, so the semantics are straightforward: a read or write is allowed after a delegation and until its revocation.
Our protocol preserves exactly these semantics whenever there is no concurrency.
In a decentralized system, however, updates can be concurrent with delegations and revocations, so we must additionally decide how a revocation affects reads and writes of updates that are concurrent with it.
We answer this question for reads in \autoref{sec:read-semantics} and for writes in \autoref{sec:authorizing-deltas}.

\subsubsection{Enforcing Read Access}
\label{sec:read-semantics}

The \RuleName*{Send Delta} and \RuleName*{Receive Delta} transition rules below govern the peer-to-peer dissemination of application state.
\RuleName*{Send Delta} determines whether the acting replica~$\eid$ may reveal a delta~$\delta$ from its store~$\Delta$
to a peer~$\eid_r$ by putting it into the peer's inbox~$I_r$; this is where read access is enforced.
\RuleName*{Receive Delta} moves a delta from the inbox of~$\eid$ into its store, provided that $\eid$ itself may read it
according to its own event graph; unsolicited deltas pushed by Byzantine peers are thus never stored.
Whether a stored delta affects the application state is additionally filtered by
materialization (\autoref{sec:materialized-state}).
This is how replicas propagate authorized payload deltas ($\Delta$)
to reconcile disparate local application state.

\begin{flushleft}
$\inferrule*[left={\makebox[6.5em][l]{\RuleName*{Send Delta}}}]{
  \delta \in \Delta
  \\ \exists e_u \in E :\;
     \mayread(E,\, \eid_r,\, e_u,\, \delta)
}{
  K[\eid \mapsto \langle E,\, \Delta,\, I \rangle,\;
    \eid_r \mapsto \langle E_r,\, \Delta_r,\, I_r \rangle]
  \;\xrightarrow{\eid}\;
  K[\eid_r \mapsto \langle E_r,\, \Delta_r,\, I_r \cup \{\delta\} \rangle]
}$
\\[3ex]
$\inferrule*[left={\makebox[6.5em][l]{\RuleName*{Receive Delta}}}]{
  \delta \in I
  \\ \exists e_u \in E :\;
     \mayread(E,\, \eid,\, e_u,\, \delta)
}{
  K[\eid \mapsto \langle E,\, \Delta,\, I \rangle]
  \;\xrightarrow{\eid}\;
  K[\eid \mapsto \langle E,\, \Delta \cup \{\delta\},\, I \rangle]
}$
\end{flushleft}

Unlike $\RuleName*{Send Event}$, which sends authorization metadata without access-control filtering because
such metadata reveals no confidential application data to unauthorized replicas, the selected $\delta$ in the rule \RuleName*{Send Delta} is filtered before it is sent:
a correct replica transmits~$\delta$ to peer~$\eid_r$ only if the predicate $\mayread$ defined in \autoref{fig:read-predicate} holds.
This filtering happens at transmission time, because a delta cannot be revoked once delivered.

Condition~(i) of $\mayread$ requires that the update event~$e_u$ commits to the delta $\delta$ being sent, bound to the author of~$e_u$.
The remaining conjuncts require some capability $e_d$ in the local event graph~$E$ that authorizes the transmission:
(ii)~the capability has a valid delegation chain,
(iii)~it is issued to the reading replica $\eid$,
(iv)~no revocation of it, or of any delegation in its chain, causally precedes~$e_u$, and
(v)~the path that $\delta$ mutates is covered by either the read or the write permissions of $e_d$.
Write permissions suffice here because modifying a value requires observing it, so write access implies read access.

Condition~(iv) is what makes read access \emph{monotonic}: a delta may still be delivered as long as the capability was not revoked at the time when the update was performed.
Updates generated while a peer held legitimate read access thus remain accessible to that peer, even if the capability is revoked concurrently or afterward.
More precisely, given an update~$e_u$ committing to~$\delta$, a capability~$e_d$ issued to~$\eid$, and a revocation $e_r \in \textit{revocations}(E, e_d)$, i.e., an authorized revocation of $e_d$ or of any delegation in its chain:
\begin{enumerate}
  \item if $e_u \prec e_r$, then reading $\delta$ via~$e_d$ remains authorized;
  \item if $e_u \parallel e_r$, then reading $\delta$ via~$e_d$ remains authorized;
  \item if $e_r \prec e_u$, then reading $\delta$ via~$e_d$ becomes unauthorized.
\end{enumerate}

\begin{figure}
\small
\begin{gather*}
\begin{aligned}
\mayread(E, \eid, e_u, \delta) :=\;
  & e_u.\payload = \Update{\hashof{(e_u.\eauthor\,\|\,\delta)}} && \text{(i)} \\
  & \wedge\; \exists e_d \in E :\; \bigl( \delegationvalid(e_d) && \text{(ii)} \\
  & \phantom{\wedge\; \exists e_d \in E :\;} \wedge\; e_d.\payload.\textit{delegatee} = \eid && \text{(iii)} \\
  & \phantom{\wedge\; \exists e_d \in E :\;} \wedge\; \neg\exists e_r \in \textit{revocations}(E, e_d) : e_r \prec e_u && \text{(iv)} \\
  & \phantom{\wedge\; \exists e_d \in E :\;} \wedge\; \epath(\delta) \sqsubseteq \bigl( e_d.\payload.\Prd \cup e_d.\payload.\Pwr \bigr) \bigr) && \text{(v)}
\end{aligned}
\end{gather*}
\caption{The predicate $\mayread$ governs read access.}
\label{fig:read-predicate}
\end{figure}

This formulation aligns with the \emph{protection relation} in weakly consistent access-control systems~\cite{weberAccessControlWeakly2016}:
all preceding revocations must be considered.
Cases~(1) and~(2) both satisfy this protection relation, since $e_r$ does not causally precede $e_u$ in either; only
case~(3), where the revocation causally precedes the commit ($e_r \prec e_u$), violates it and strips read access.
Peers thus retain the right to read deltas committed during the valid epoch of their capability, and only causally
preceding revocations remove that right.

Enforcing a stricter policy against concurrent reads requires a total causal serialization. Our system can achieve this by
forcing all communication through a centralized coordinator~\cite{dougalERAEpochResolvedArbitration2026}, at the cost of giving up the benefits of decentralization.
This shows that our solution is a superset of centralized enforcement: it recovers the same guarantees whenever that
stricter setting is imposed.

\subsubsection{Enforcing Write Access}
\label{sec:authorizing-deltas}
\label{sec:materialized-state}

Write access is enforced locally by each replica when it decides which of the deltas in its delta store~$\Delta$ are \emph{included} in its application state (\autoref{fig:write-predicate}).
Write authority is not checked on receipt and may change: a Byzantine peer could send deltas whose authors lack write permissions, or deltas authorized by capabilities whose concurrent revocation arrives only later.
The \emph{materialized} application state is the merge~$\bigsqcup$ of all included deltas.

A delta~$\delta$ is included if some update event~$e_u$ committing to it satisfies the predicate $\maywrite$, which requires that:
(i)~the event commits to the delta~$\delta$, bound to the author of the event,
(ii)~the path modified by $\delta$ is covered ($\sqsubseteq$) by the write permissions of the authorizing capability $e_u.\eauth$,
(iii)~the author of $e_u$ is the designated delegatee of that capability,
(iv)~every known revocation of the authorizing capability, or of any delegation in its chain, causally follows the event, i.e., is neither concurrent with nor before it, and
(v)~the authorizing capability is supported by a valid delegation chain.

Condition~(iv) implements a \emph{revocation-wins} semantics: writes that are concurrent with a revocation of their authorizing capability are rejected.
More precisely, given an update event~$e_u$ and a revocation $e_r \in \textit{revocations}(E, e_u.\eauth)$, i.e., an authorized revocation of $e_u.\eauth$ or of any delegation in its chain:
\begin{enumerate}
  \item if $e_u \prec e_r$, then $e_u$ remains authorized;
  \item if $e_u \parallel e_r$, then $e_u$ becomes unauthorized;
  \item if $e_r \prec e_u$, then $e_u$ becomes unauthorized.
\end{enumerate}
This semantics eliminates the \emph{backdating loophole}. In a Byzantine setting, an adversarial replica can deliberately omit $e_r$ from its predecessor set and thus manufacture an apparent concurrency ($e_u \parallel e_r$). These \emph{backdated} writes are still treated as unauthorized.

\begin{figure}
\small
\begin{gather*}
\begin{aligned}
\maywrite(E, e_u, \delta) :=\;
           & e_u.\payload = \Update{\hashof{(e_u.\eauthor\,\|\,\delta)}} && \text{(i)} \\
  \wedge\; & \epath(\delta) \sqsubseteq e_u.\eauth.\payload.\Pwr && \text{(ii)} \\
  \wedge\; & e_u.\eauth.\payload.\textit{delegatee} = e_u.\eauthor && \text{(iii)} \\
  \wedge\; & \forall e_r \in \textit{revocations}(E,\, e_u.\eauth) : e_u \prec e_r && \text{(iv)} \\
  \wedge\; & \delegationvalid(e_u.\eauth) && \text{(v)} \\[1ex]
\mathit{included}(E,\, \Delta)
\;:=\; &
\{\, \delta \in \Delta \mid \exists e_u \in E :\;
\maywrite(E, e_u, \delta) \,\} \\
\mathit{materialized}(E,\, \Delta)
\;:=\; &
\bigsqcup \mathit{included}(E,\, \Delta)
\end{aligned}
\end{gather*}
\caption{Write access: the predicate $\maywrite$ and state materialization.}
\label{fig:write-predicate}
\label{fig:materialization}
\end{figure}

\section{Correctness}
\label{sec:correctness}

In this section, we prove that our protocol provides three access-control guarantees:
\emph{integrity}, i.e., only authorized writes affect the materialized state (Section~\ref{sec:correctness-write});
\emph{confidentiality}, i.e., correct replicas never disclose deltas to replicas that may not read them (Section~\ref{sec:correctness-read});
and \emph{availability}, i.e., every authorized write eventually reaches every correct replica that may read it (Section~\ref{sec:correctness-convergence}).
These results build on basic properties of the event graph (Section~\ref{sec:correctness-graph}) and on the fact that delegations cannot expand permissions (Section~\ref{sec:correctness-no-expansion}).
We further show that, without concurrent revocations, our semantics match those of a centralized system (Section~\ref{sec:correctness-total-order}).

All guarantees hold in the presence of Byzantine replicas (Section~\ref{sec:system-model}).
Byzantine replicas may put arbitrary events and deltas into the inbox of any other replica, and they may read everything in their own inbox.
They are, however, bound by the cryptographic assumptions of our threat model: they cannot forge signatures of other replicas, find collisions, or invert hashes.

Our goals are therefore twofold.
First, no confidential delta may reach the inbox of a Byzantine replica that lacks read access to it.
Second, whatever Byzantine replicas place into the inbox of a correct replica must not cause it to include an unauthorized write in its materialized state.
Purely local operations are irrelevant to access control, so we do not further consider the \RuleName*{Action} rule in the proofs, and only assume that correct replicas do not make up data for paths they may not read.

\subsection{Event Graph Properties}
\label{sec:correctness-graph}
\label{sec:correctness-delegation}

All subsequent results rely on a few basic properties of event graphs.
Lemma~\ref{lem:dag-acyclic} is less a result in its own right than a summary of what our assumptions provide: the properties follow almost directly from the cryptographic assumptions of our threat model, the construction of events in \RuleName*{Action}, and the $\causallyready$ check in \RuleName*{Receive Event}.

\begin{lemma}[Well-Formed Event Graphs]
\label{lem:dag-acyclic}
Let $E$ be the event graph of a correct replica in a reachable configuration. Then:
\begin{enumerate}
  \item the happened-before relation $\prec$ (Definition~\ref{def:event-graph}) is a strict partial order on~$E$, and $E$ contains all causal predecessors of its events;
  \item the genesis event $e_\bot$ is the unique minimal event of~$E$;
  \item the \epre and \eauth links of an event cannot be changed or forged;
  \item for every $e \in E$ with $e \neq e_\bot$, we have $e.\eauth \prec e$, and thus $\eauth^*(e) \subseteq E$.
\end{enumerate}
\end{lemma}

\begin{proof}[Proof sketch]
$E$ starts as $\{e_\bot\}$ and only grows by \RuleName*{Action} and \RuleName*{Receive Event}.
\RuleName*{Action} creates an event whose predecessors are the (non-empty) heads of~$E$ and whose authorizing capability is in~$E$;
\RuleName*{Receive Event} admits only causally ready events, whose predecessors are in~$E$ and, unless the event is $e_\bot$ (which is already in~$E$), non-empty and include the authorizing capability in their causal past (Section~\ref{sec:action-rule}).
Thus every new event only points to existing events, which yields (1), (2), and (4).
(3)~follows from the cryptographic assumptions (Section~\ref{sec:system-model}): links are collision-resistant hashes inside a signed event.
\end{proof}

Since the event graph is grow-only and replicated without access-control filtering, it converges under the eventual-delivery assumption (Section~\ref{sec:system-model}).

\begin{theorem}[Event Graph Convergence]
\label{thm:event-convergence}
Every event in the event graph of a correct replica is eventually in the event graph of every correct replica.
\end{theorem}

\begin{proof}
Event graphs are grow-only sets, and \RuleName*{Send Event} and \RuleName*{Receive Event} filter nothing.
The only premise, $\causallyready$, merely orders deliveries: by Lemma~\ref{lem:dag-acyclic}, the predecessors of an event of a correct replica are also in its graph and are thus delivered as well.
\end{proof}

\subsection{Capabilities Cannot Be Expanded}
\label{sec:correctness-no-expansion}

We prove that colluding Byzantine replicas cannot extend their permissions.
Intuitively, whatever permissions a group of Byzantine replicas delegates, they are no more than the permissions delegated to the group from the outside, i.e., by a correct replica.

\begin{theorem}[No Expansion of Permissions]
\label{thm:no-expansion}
Let $\mathcal{B} \subseteq \mathbb{I}$ be a set of (possibly colluding) Byzantine replicas,
and let $e_d$ be a delegation event with $\mathit{delegation\_valid}(e_d)$ and $e_d.\eauthor \in \mathcal{B}$.
Then there is a delegation event $e_g \in \eauth^*(e_d)$ that is either $e_\bot$ or authored by a
correct replica, that delegates to a member of the group ($e_g.\payload.\textit{delegatee} \in \mathcal{B}$),
and $e_d$ grants no more read and write permissions than $e_g$
($e_d.\payload.\Prd \sqsubseteq e_g.\payload.\Prd$ and $e_d.\payload.\Pwr \sqsubseteq e_g.\payload.\Pwr$).
\end{theorem}

\begin{proof}
If the initiating replica $\eid_\bot = e_\bot.\payload.\textit{delegatee}$ is in~$\mathcal{B}$, the Byzantine replicas already hold full permissions via~$e_\bot$, which trivially cannot be expanded further.
Otherwise, $\eid_\bot \notin \mathcal{B}$.
Since $\mathit{delegation\_valid}(e_d)$ holds, unfolding its recursive clause shows that every delegation in the
authorization chain $\eauth^*(e_d)$ satisfies \delegationvalid, and that the chain contains~$e_\bot$, the
only base case.
Every event is signed by its author (Definition~\ref{def:event}), so an event whose author is a correct replica was
indeed created by that replica: replicas in~$\mathcal{B}$ cannot sign in its name.

Follow the \eauth pointers from $e_d$ towards $e_\bot$, and let $e_g$ be the first delegation reached that is
$e_\bot$ or authored by a correct replica.
Let $e_h$ be the delegation visited just before $e_g$, i.e., $e_h.\eauth = e_g$.
All delegations visited from $e_d$ up to $e_h$ are authored by replicas in~$\mathcal{B}$, and by $\delegationvalid(e_h)$, the author of $e_h$ is the delegatee of its parent, so
$e_g.\payload.\textit{delegatee} = e_h.\eauthor \in \mathcal{B}$.
Thus, $e_g$ is the delegation by which a correct replica initially granted permissions to the group~$\mathcal{B}$.
By $\delegationvalid$, each of these delegations grants a subset of the permissions of its parent;
since $\sqsubseteq$ is transitive, $e_d$ grants a subset of the permissions of~$e_g$, i.e., the permissions initially granted by the correct replica have not been expanded within the group.
\end{proof}

\subsection{Write Integrity}
\label{sec:correctness-write}

Write integrity states that every write needs to be justified by a non-revoked delegation.
Because later revocations may invalidate already received writes, we enforce integrity by filtering the deltas included by materialization (Figure~\ref{fig:materialization}).

\begin{theorem}[Write Integrity]
\label{thm:write-integrity}
Let $E$ be an event graph and $\Delta$ a delta store.
If $\delta \in \mathit{included}(E, \Delta)$, then
\begin{enumerate}
  \item there is an update event $e_u \in E$ that commits to~$\delta$, i.e., $e_u.\payload = \Update{\hashof{(e_u.\eauthor\,\|\,\delta)}}$,
        and that is authorized by a delegation $e_d = e_u.\eauth$ that grants $e_u.\eauthor$ write access to
        $\epath(\delta)$,
  \item that is not revoked, i.e., for every $e \in \eauth^*(e_d)$ and every $e_r \in E$ with $e_r.\payload = \Revoke{\hashof{e}}$ and
        $\mayrevoke(e_r)$, we have $e_u \prec e_r$.
\end{enumerate}
\end{theorem}

\begin{proof}
By definition of $\mathit{included}$, there is an $e_u \in E$ with $\maywrite(E, e_u, \delta)$.
(1)~is conjuncts (i)--(iii) of $\maywrite$ (Figure~\ref{fig:write-predicate}).
(2)~$\mathit{revocations}(E, e_d)$ recurses along the \eauth pointers
(Figure~\ref{fig:delegation-validity-revocation-predicates}), so it contains every $e_r$ described in~(2).
Conjunct (iv) of $\maywrite$ requires $e_u \prec e_r$ for all of them.
\end{proof}

Theorem~\ref{thm:write-integrity} only states that a non-revoked delegation~$e_d$ exists.
Since $\maywrite$ also requires $\mathit{delegation\_valid}(e_d)$, Theorem~\ref{thm:no-expansion} additionally
ensures that $e_d$ stems from a delegation chain that does not expand rights: even if Byzantine replicas created
$e_d$, it grants no more than they were given by a correct replica or the genesis event.

\subsection{Read Confidentiality}
\label{sec:correctness-read}
We prove that correct replicas do not reveal confidential writes%
\footnote{We cannot guarantee confidentiality of data that Byzantine replicas may read:
they can simply send it to unauthorized replicas, or, with sufficient write access, re-commit it in a backdated update that is concurrent with a revocation of another replica's read access, making it readable by the revoked replica.
Such leaks are outside the scope of our guarantees.}, i.e., delta values,
to replicas lacking the corresponding read permissions.
A delta is revealed to a replica as soon as it is placed into that replica's inbox, which for correct senders only
happens via \RuleName*{Send Delta}; whether and when the receiver $\eid_r$ takes it out via \RuleName*{Receive Delta} is
irrelevant for confidentiality.
Theorem~\ref{thm:read-filtering} shows that a correct sender places a delta~$\delta$ into the inbox of
$\eid_r$ only when $\eid_r$ holds a valid read capability for it, and that this decision is never
invalidated by revocations that arrive later.

The decision whether a delta may be sent must not be reversed by events that arrive later.
Read access for a specific delta is indeed never revoked:

\begin{lemma}[Read Authorization Is Stable]
\label{lem:read-stable}
Let $E$ and $E'$ with $E \subseteq E'$ be causally closed event graphs.
If $e_u \in E$ and $\mayread(E, \eid, e_u, \delta)$, then $\mayread(E', \eid, e_u, \delta)$.
\end{lemma}

\begin{proof}
Take the delegation $e_d$ chosen by $\mayread(E, \eid, e_u, \delta)$, which is in $E'$ since $E \subseteq E'$. No conjunct of $\mayread$ except the
revocation conjunct depends on the event graph, so it remains to show that no $e_r \in \mathit{revocations}(E', e_d)$ satisfies $e_r \prec e_u$.
Suppose such an $e_r$ exists. Then $e_r$ is a causal predecessor of $e_u \in E$, so $e_r \in E$.
As $\mayrevoke$ does not depend on the event graph, $e_r \in \mathit{revocations}(E, e_d)$, contradicting
$\mayread(E, \eid, e_u, \delta)$.
\end{proof}

Correct replicas' event graphs satisfy the premise, since \RuleName*{Receive Event} only admits causally ready events.

\begin{theorem}[Read Confidentiality]
\label{thm:read-filtering}
Let a correct replica $\eid$ with event graph $E$ take a \RuleName*{Send Delta} step that places~$\delta$
into the inbox of $\eid_r$.
Then there is an update event $e_u \in E$ that commits to~$\delta$ and a delegation $e_d \in E$ with
$\delegationvalid(e_d)$ that grants $\eid_r$ read (or write) access to $\epath(\delta)$,
and this send is never invalidated: $\mayread(E', \eid_r, e_u, \delta)$ holds for every later event graph $E'$
of any correct replica with $E \subseteq E'$.
\end{theorem}

\begin{proof}
Since $\eid$ is correct, the premise of \RuleName*{Send Delta} holds:
$\mayread(E, \eid_r, e_u, \delta)$ for some $e_u \in E$, whose conjuncts
(Figure~\ref{fig:read-predicate}) give the delegation~$e_d$.
By Lemma~\ref{lem:read-stable}, $\mayread$ still holds in~$E'$.
\end{proof}

Together with Theorem~\ref{thm:no-expansion}, this ensures that the path is legitimately readable by
$\eid_r$: $e_d$ stems from a delegation chain that does not expand rights.
Intuitively, every revocation that causally precedes $e_u$ is already known to the sender, %
and revocations that are concurrent with
or causally after $e_u$ do not apply to reads (Section~\ref{sec:read-semantics}).
A revocation $e_r \parallel e_u$ may be unknown to any replica holding~$\delta$, in particular to the author of~$e_u$,
which may already have sent~$\delta$ before learning of~$e_r$; a delta, once placed into an inbox, cannot be retracted.
This is precisely why concurrent revocations do not remove read access: it would not be possible to enforce.
Enforcing concurrent revocations would make the read policy depend on delivery order, which no correct replica can guarantee.
Revocations still take effect for causally dependent updates: any update event~$e_u'$ created after a replica learned of~$e_r$ is causally
after~$e_r$, so the revocation applies to~$e_u'$.
No correct replica sends the delta for $e_u'$ based on a delegation that was revoked causally before $e_u'$.

\begin{theorem}[Read Confidentiality After Revocation]
\label{thm:read-revocation}
Let a correct replica \eid with event graph~$E$ create an update event~$e_u$ committing to~$\delta$, and let $e_d$ be a delegation that is revoked in~$E$, i.e., $\mathit{revocations}(E, e_d) \neq \emptyset$.
Then no correct replica ever sends~$\delta$ to $\eid_r = e_d.\payload.\textit{delegatee}$ based on~$e_d$.
\end{theorem}

\begin{proof}
A correct replica with event graph~$E'$ places~$\delta$ into the inbox of~$\eid_r$ only via
\RuleName*{Send Delta}, which requires some $e_u' \in E'$ with $\mayread(E', \eid_r, e_u', \delta)$ that in turn requires a delegation $e_d'$ that is not revoked in $E$.
We show that $e_u$ is the only possible update $e_u' = e_u$ and that it was authorized by a non-revoked delegation $e_d' \neq e_d$.

\emph{$e_u' = e_u$.}
$\mayread$ requires that $e_u'$ commits to~$\delta$ written by $e_u'.\eauthor$.
Since events are signed, $e_u'$ was created by $e_u'.\eauthor$.
If $e_u'.\eauthor = \eid$, i.e., the correct replica that created~$e_u$, then because of the fresh salt (\autoref{sec:commitments}), it commits to~$\delta$ only once, so $e_u' = e_u$.
If $e_u'.\eauthor \neq \eid$, the salt in $\delta$ is not fresh, so $e_u'.\eauthor$ is Byzantine; we thus disregard this case.

\emph{$e_d' \neq e_d$.}
The predecessors of $e_u$ are the heads of~$E$, so $e \prec e_u$ for every $e \in E$. As $e_u \in E'$ and $E'$ is
closed under predecessors (Lemma~\ref{lem:dag-acyclic}), $E \subseteq E'$.
Now let $e_d'$ be the delegation chosen by $\mayread(E', \eid_r, e_u, \delta)$.
By definition of \mayread, there is no revocation $e_r \in \textit{revocations}(E', e_d') : e_r \prec e_u$.
Thus, if such an $e_r$ is not in $E'$, it also cannot be in $E$, as $\textit{revocations}(E, e_d') \subseteq \textit{revocations}(E', e_d')$ with $E \subseteq E'$.
By assumption $e_d$ has a revocation $e_r \in E$, which implies that $e_d' \neq e_d$.
\end{proof}

\paragraph{Receiving.}
A Byzantine replica could simply send deltas to a correct replica that the latter is not permitted to read.
This is technically no breach of confidentiality, since information from Byzantine replicas is not protected.
However, it is still undesired: a correct replica should not store or process data it may not read.

\begin{corollary}[No Unauthorized Reception]
\label{cor:no-unauthorized-reception}
In any reachable configuration, the delta store~$\Delta$ of a correct replica $\eid$ with event graph $E$
(and thus its materialized state) contains no delta it may not read: for every $\delta \in \Delta$ there is an
$e_u \in E$ with $\mayread(E, \eid, e_u, \delta)$.
\end{corollary}

\begin{proof}
A delta enters $\Delta$ only through an \RuleName*{Action} step, where $\mayread$ holds by the threat-model
assumption, or through \RuleName*{Receive Delta}, whose premise excludes every delta the replica may not read.
As $E$ grows afterwards, $\mayread$ remains valid by Lemma~\ref{lem:read-stable}.
\end{proof}

\subsection{Centralized Semantics Without Concurrency}
\label{sec:correctness-total-order}

To further strengthen the case for the design decisions made in our protocol, we compare the semantics to a centralized system that rules out concurrent operations. In particular, we show that without concurrent revocations, our semantics match those of a centralized system.

\begin{corollary}[Centralized Semantics Without Concurrency]
\label{lem:total-order}
If no event in~$E$ is concurrent with a revocation, then an update~$e_u$ is authorized for reading and writing via a delegation~$e_d$ if and only if $e_d$ was not yet revoked when $e_u$ was committed, as in a centralized access-control system.
\end{corollary}

\begin{proof}
Writes before a revocation are authorized and writes after it are not (Theorem~\ref{thm:write-integrity}); the same holds for reads (Theorem~\ref{thm:read-revocation}). Without concurrency, these are the only cases.
\end{proof}

Here, a read takes place at commit time: a delta delivered after a revocation but committed before it corresponds to a read that a centralized system would have permitted before the revocation.

\subsection{Availability and Convergence}
\label{sec:correctness-convergence}

Availability, as a security goal, states that every write, i.e., every delta~$\delta$ committed by an update event~$e_u$ with $\maywrite(E, e_u, \delta)$, eventually becomes visible to every correct replica that may read it.
Since the policy evolves dynamically, this first requires that delegations and revocations are eventually applied at
every correct replica, which is guaranteed by Theorem~\ref{thm:event-convergence}.

First, Lemma~\ref{lem:write-implies-read} shows that write access implies read access, so every authorized delta has
at least one reader, its author, who includes it.
Then, Theorem~\ref{thm:availability} shows that if a correct replica includes a delta and believes that another
correct replica may read it too, then that other replica eventually includes the delta as well.

\begin{lemma}[Write Authority Implies Read Authority]
\label{lem:write-implies-read}
For every event graph~$E$ of a correct replica in a reachable configuration, update event~$e_u \in E$, and delta~$\delta$:
\[
  \maywrite(E, e_u, \delta) \implies \mayread(E, e_u.\eauthor, e_u, \delta).
\]
\end{lemma}

\begin{proof}
The predicate $\mayread$ (Figure~\ref{fig:read-predicate}) asks for \emph{some} delegation $e_d$ that grants
the reader access to the path, and it accepts write permissions as well as read permissions for this.
$\maywrite$ already provides such a capability: the delegation $e_d = e_u.\eauth$ that authorizes the write.
We check that it satisfies each conjunct of $\mayread$:
\begin{itemize}
  \item $e_d \in E$: by Lemma~\ref{lem:dag-acyclic}~(4), the authorizing capability of $e_u \in E$ is in~$E$.
  \item $e_u$ commits to~$\delta$ and $e_d$ is valid: both are also conjuncts of $\maywrite$.
  \item $e_d$ is granted to the reader $e_u.\eauthor$\,: $\maywrite$ requires $e_u.\eauthor$ to be the delegatee of~$e_d$.
  \item $e_d$ covers the path: $\maywrite$ gives $\epath(\delta) \sqsubseteq e_d.\payload.\Pwr$, which is part of
        $e_d.\payload.\Prd \cup e_d.\payload.\Pwr$.
  \item $e_d$ is not revoked before $e_u$: $\maywrite$ requires every revocation of $e_d$ to come after~$e_u$, and an
        event cannot be both before and after~$e_u$.
\end{itemize}
\end{proof}

\begin{theorem}[Availability]
\label{thm:availability}
Let $\eid$ be a correct replica with event graph $E$ and delta store $\Delta$ that includes a delta~$\delta$,
with the corresponding update event $e_u$ with $\maywrite(E, e_u, \delta)$, and let $\eid_r$ be a correct replica with
$\mayread(E, \eid_r, e_u, \delta)$.
Then eventually, for the event graph $E'$ and delta store $\Delta'$ of $\eid_r$, either
$\delta \in \mathit{included}(E', \Delta')$, or
$E'$ contains a revocation $e_r \in \mathit{revocations}(E', e_u.\eauth)$ with $e_u \not\prec e_r$.
\end{theorem}

\begin{proof}
By Theorem~\ref{thm:event-convergence}, the event graph of $\eid_r$ eventually contains~$E$.
As $\mayread$ remains valid when the event graph grows (Lemma~\ref{lem:read-stable}),
$\mayread(\cdot, \eid_r, e_u, \delta)$ then holds in the graphs of both $\eid$ and $\eid_r$ and keeps
holding: read access for a specific delta is never revoked (cf.\ Theorem~\ref{thm:read-filtering}).
Thus \RuleName*{Send Delta} of~$\delta$ from $\eid$ to $\eid_r$ and then \RuleName*{Receive Delta} at
$\eid_r$ are enabled and remain so, and by eventual delivery $\delta \in \Delta$ eventually.
Deltas are never removed, so $\delta$ is included if
$\maywrite(E', e_u, \delta)$.
All conjuncts of $\maywrite$ except the revocation conjunct do not depend on the event graph and hold since they
hold in~$E$; so if $\maywrite(E', e_u, \delta)$ fails, there is an $e_r \in \mathit{revocations}(E', e_u.\eauth)$ with
$e_u \not\prec e_r$.
\end{proof}

Beyond availability, access control also preserves the convergence of the underlying ARDTs among replicas with
the same read permissions:

\begin{corollary}[Equal Read Access, Equal State]
\label{cor:sec}
Let $\eid_1$ and $\eid_2$ be correct replicas, and suppose that from some point on no new events are
created, so that both eventually hold the same event graph~$E$ (Theorem~\ref{thm:event-convergence}).
If $\mayread(E, \eid_1, e_u, \delta) \Leftrightarrow \mayread(E, \eid_2, e_u, \delta)$ for all $e_u$
and~$\delta$, then eventually their materialized states are equal.
\end{corollary}

\begin{proof}
By Theorem~\ref{thm:availability}, every delta included at one replica is eventually included at the other, as both
share~$E$ and agree on $\mayread$; so the included sets eventually coincide.
The materialized states are then joins of the same deltas, which are equal as the ARDT merge is associative,
commutative, and idempotent.
\end{proof}

\section{Evaluation}\label{chap:eval}

We empirically quantify the computational overhead of the GUARDTs event graph relative to two HashDAGs without access control as baselines, with and without signatures.

\subsection{Experimental Setup}

\paragraph{System Variants}

HashDAGs are a common construction for Byzantine fault-tolerant CRDTs~\cite{kleppmannByzantineEventualConsistency2020, jacobCRDTsByzantineEnvironments2022, kleppmannMakingCRDTsByzantine2022}. They are used in Matrix~\cite{jacobAnalysisMatrixEvent2021} and also in Automerge, which is not Byzantine fault-tolerant.
We therefore compare GUARDTs against two HashDAG baselines without access control, in which each entry records the \gls{ardt} delta, the author's identity, and the hashes of the preceding heads.
The baselines differ only in whether entries are signed with Ed25519.
Signatures are needed in any Byzantine setting to authenticate authors, independently of access control.
Comparing against both baselines thus separates the cost of signatures from the additional cost of our access-control mechanism.

The overhead of access control does not depend on the semantics of the replicated application state, i.e., on how deltas are merged.
We therefore choose a cheap application \gls{ardt} to isolate the access-control overhead from the cost of the application itself.
For more expensive \glspl{ardt}, the absolute overhead stays the same and its relative share only shrinks.
Concretely, we use nested product \glspl{ardt} of counters: the products provide multiple paths to exercise access control, while counters have a small, fixed cost per operation.
We name the fields \texttt{a}, \texttt{b}, and \texttt{c} at each level, so a path to a counter is, e.g., \texttt{a.a.b}.

\paragraph{Benchmarks}

We benchmark the isolated overhead of individual event graph operations:
\begin{enumerate}
  \item \emph{Creating Updates}: creating an event, computing the 
  delta and its commitment, and selecting a capability with 
  sufficient permissions;
  \item \emph{Sending Updates} to a peer: 
  filtering according to the peer's read permissions;
  \item \emph{Receiving Updates}: integrating the event into the local 
  event graph, and merging the delta into the materialized application state;
  \item \emph{Applying Revocations}: adding the revocation event to the graph and, if necessary, invalidating updates that become unauthorized and recomputing the materialized state.
\end{enumerate}

\colorlet{delegateText}{delegateColor!60!black}
\colorlet{revokeText}{revokeColor!60!black}
\colorlet{updateText}{black}

\newcommand{\trevsize}{6mm}
\newcommand{\trevnode}[4]{%
  \node[#1] (#2) at #3 {};%
  \node[trevlabel, text=#1Text] at (#2) {#4};%
}

\begin{figure}
\centering
\begin{tikzpicture}[
  x=1.4cm, y=1cm,
  replica/.style={font=\scriptsize, align=center, inner sep=1pt},
  delegation/.style={-latex, thick, draw=delegateColor,
                     shorten >=1pt, shorten <=1pt},
  ]
  \node[replica] (r0) at (0,0) {$\mathit{id}_\bot$: \texttt{*}};
  \foreach \i/\p [count=\k from 0] in {1/a, 2/b, 3/c}{%
    \node[replica] (r\i) at (3*\k-3,-1) {$\mathit{id}_{\i}$: \texttt{\p}};
    \draw[delegation] (r0) -- (r\i);
    \foreach \q [count=\l from 0] in {a, b, c}{%
      \pgfmathtruncatemacro{\j}{3*\i+\l+1}
      \node[replica] (r\j) at (3*\k+\l-4,-2) {$\mathit{id}_{\j}$: \texttt{\p.\q}};
      \draw[delegation] (r\i) -- (r\j);
    }
  }
\end{tikzpicture}

\bigskip
\begin{tikzpicture}[
  x=0.85cm, y=1cm,
  event/.style={draw, thick, inner sep=0pt, minimum size=\trevsize},
  delegate/.style={event, diamond, aspect=1, draw=delegateColor, fill=delegateColor!15},
  revoke/.style={event, regular polygon, regular polygon sides=8,
                 draw=revokeColor, fill=revokeColor!15},
  update/.style={event, circle, draw=updateColor, fill=none},
  trevlabel/.style={font=\scriptsize, inner sep=0pt},
  elided/.style={inner sep=1pt, minimum size=4mm},
  pre/.style={-latex, thick, shorten >=1pt, shorten <=1pt},
  path/.style={font=\scriptsize, text height=1.6ex, text depth=0.4ex, text=delegateText, inner sep=1pt},
  target/.style={font=\scriptsize, text height=1.6ex, text depth=0.4ex, text=revokeText, inner sep=1pt},
  uname/.style={font=\scriptsize, text height=1.6ex, text depth=0.4ex, text=updateText, inner sep=1pt},
  phase/.style={decorate, decoration={brace, amplitude=4pt, mirror}, thick, gray},
  ]
  \trevnode{delegate}{g} {(0,0)}{$\mathit{id}_\bot$}
  \trevnode{delegate}{da}{(1,0)}{$\mathit{id}_\bot$}
  \trevnode{delegate}{db}{(2,0)}{$\mathit{id}_\bot$}
  \trevnode{delegate}{dc}{(3,0)}{$\mathit{id}_\bot$}
  \trevnode{update}{u1}  {(4,0)}{$\mathit{id}_3$}
  \node[elided] (e1) at (5,0) {$\cdots$};
  \trevnode{update}{u2}  {(6,0)}{$\mathit{id}_2$}

  \trevnode{delegate}{daa}{(7,0)}{$\mathit{id}_1$}
  \node[elided] (e2) at (8,0) {$\cdots$};
  \trevnode{delegate}{dcc}{(9,0)}{$\mathit{id}_3$}
  \trevnode{update}{u3}  {(10,0)}{$\mathit{id}_2$}
  \node[elided] (e3) at (11,0) {$\cdots$};
  \trevnode{update}{u5}  {(12,0)}{$\mathit{id}_9$}
  \trevnode{update}{u6}  {(13,0)}{$\mathit{id}_4$}

  \node[path, below=1pt of g]   {$\mathit{id}_\bot$: \texttt{*}};
  \node[path, below=1pt of da]  {$\mathit{id}_1$: \texttt{a}};
  \node[path, below=1pt of db]  {$\mathit{id}_2$: \texttt{b}};
  \node[path, below=1pt of dc]  {$\mathit{id}_3$: \texttt{c}};
  \node[path, below=1pt of daa] {$\mathit{id}_4$: \texttt{a.a}};
  \node[path, below=1pt of dcc] {$\mathit{id}_{12}$: \texttt{c.c}};

  \node[uname, below=1pt of u1] {$u_1$};
  \node[uname, below=1pt of u2] {$u_{n/2}$};
  \node[uname, below=1pt of u3, anchor=north west, xshift=-\trevsize/2] {$u_{(n/2)+1}$};
  \node[uname, below=1pt of u5] {$u_{n-1}$};
  \node[uname, below=1pt of u6] {$u_n$};

  \foreach \later/\earlier in {da/g, db/da, dc/db, u1/dc, e1/u1, u2/e1,
      daa/u2, e2/daa, dcc/e2, u3/dcc, e3/u3, u5/e3, u6/u5}{%
    \draw[pre] (\later) -- (\earlier);
  }

  \trevnode{revoke}{ra}  {(2,1)}{$\mathit{id}_\bot$}
  \trevnode{revoke}{raa} {(8,1)}{$\mathit{id}_1$}
  \trevnode{revoke}{raa2}{(13,1)}{$\mathit{id}_1$}
  \trevnode{revoke}{rna} {(14,0)}{$\mathit{id}_\bot$}
  \draw[pre] (ra)   -- (da);
  \draw[pre] (raa)  -- (daa);
  \draw[pre] (raa2) -- (u5);
  \draw[pre] (rna)  -- (u6);
  \node[target, right=1pt of ra]   {$\mathit{rev}_1$: $\mathit{id}_1 - {}$\texttt{a}};
  \node[target, right=1pt of raa]  {$\mathit{rev}_2$: $\mathit{id}_4 - {}$\texttt{a.a}};
  \node[target, right=1pt of raa2] {$\mathit{rev}_3$: $\mathit{id}_4 - {}$\texttt{a.a}};
  \node[target, below=1pt of rna, xshift=1mm]  {$\mathit{rev}_4$: $\mathit{id}_1 - {}$\texttt{a}};

  \draw[phase] (-0.3,-0.75) -- node[below=4pt, font=\scriptsize, text=black]
    {phase 1: $\frac{n}{2}$ updates by $\mathit{id}_1$--$\mathit{id}_3$} (6.4,-0.75);
  \draw[phase] (6.6,-0.75) -- node[below=4pt, font=\scriptsize, text=black]
    {phase 2: $\frac{n}{2}$ updates by $\mathit{id}_1$--$\mathit{id}_{12}$} (13.4,-0.75);

  \begin{scope}[shift={(3,-1.7)}, font=\scriptsize]
    \node[delegate, scale=0.7] (lg) at (0,0) {};
    \node[right=1.5mm of lg] {delegate};
    \node[update, scale=0.7] (lu) at (2.4,0) {};
    \node[right=1.5mm of lu] {update};
    \node[revoke, scale=0.7] (lr) at (4.6,0) {};
    \node[right=1.5mm of lr] {revoke};
    \draw[pre] (7.6,0) -- (6.8,0);
    \node[anchor=west] at (7.7,0) {\epre};
  \end{scope}
\end{tikzpicture}
\caption{%
  The benchmark trace with the delegation hierarchy at the top showing the write paths
  delegated to each replica. At the bottom is the event graph of the trace, with events 
  labeled with their author. The~"$\cdots$" stand for elided events, and
  $n \in \{10\,000, 100\,000\}$ is the number of updates. Delegations are
  annotated with the replica they delegate to and the write path they delegate, and revocations with the write
  path of the capability they revoke. While all revocations are shown in the graph,
  they are only included in the revocation-specific benchmarks and only one of them per benchmark.
}
\label{fig:trace}
\end{figure}
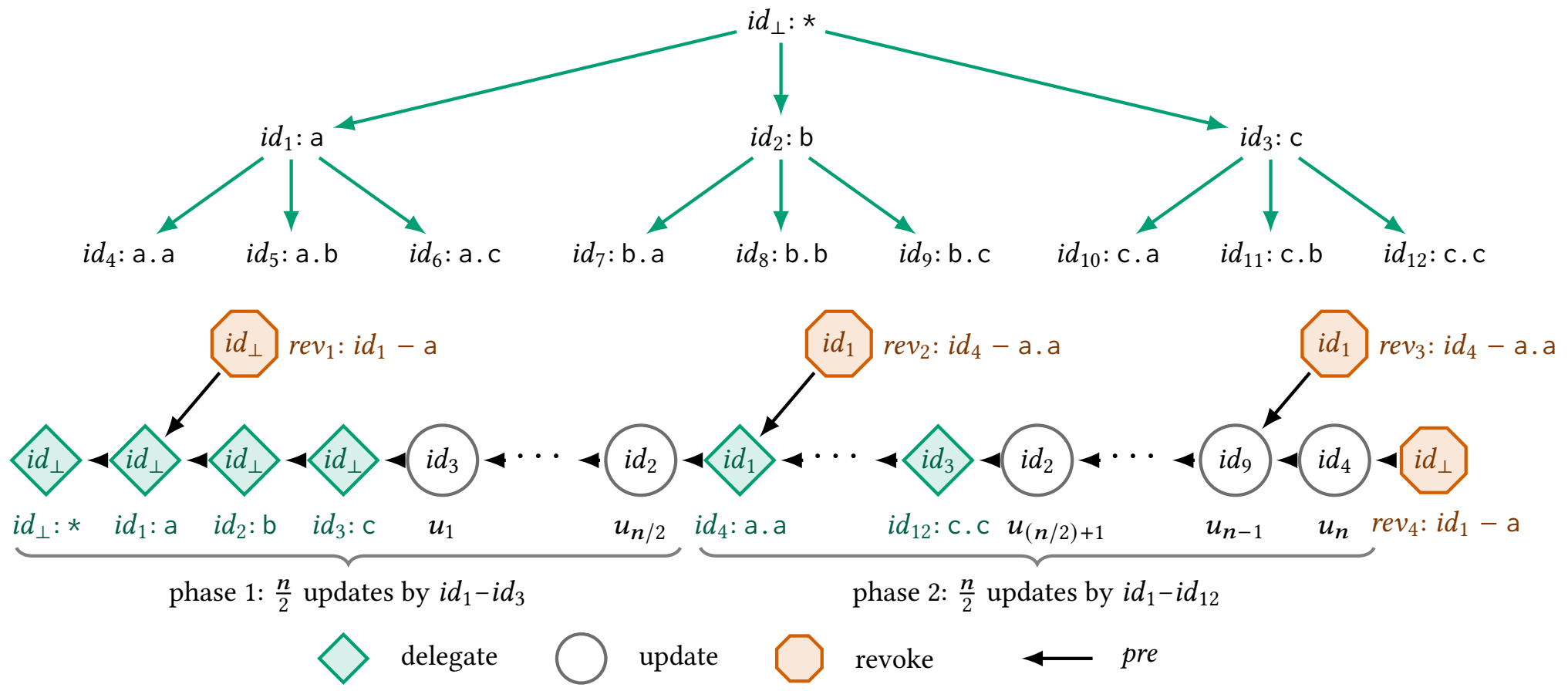

The benchmark uses a generated trace with two phases, each containing half of the updates (\autoref{fig:trace}).
In the first, an initial replica delegates partial write and full read access to three replicas, which then perform updates.
In the second, each of these replicas delegates parts of its access to three new replicas, after which all replicas perform updates.
Altogether, the trace has 13 replicas: the initial replica, which only delegates, and 12 that concurrently perform updates.
We use a trace with 100\,000 updates. Trace size has negligible impact on operations (1) to (3), so we additionally use a trace with 10\,000 updates only for revocations (4).
To ensure a fair comparison, all updates are authorized and every replica has full read access: unauthorized updates would be filtered out before reaching the application state, and withheld deltas would not be sent, both artificially reducing the cost of the system with access control.
We report the computational cost of a single replica executing each operation.
Only the author creates an update (1), while other replicas also send (2) and receive (3) it.
Operations (1) to (3) are thus on the critical path until an update becomes visible at a directly connected replica, while relaying an existing update involves only (2) and (3).

Receiving a revocation is only expensive if the receiver has already applied an update that is concurrent with the revocation and authorized by the revoked delegation.
The materialized state must then be fully recomputed without that update, because ARDT deltas have no inverse that could subtract it.
We assume that revocations are rare compared to updates, and revocations concurrent with an affected update rarer still, especially among well-connected replicas. We thus quantify the cost of recomputation but exclude it from the critical path.

For benchmark (4), the cost depends on the size of the event graph and the revocation's position within it, so we evaluate four revocation scenarios, $\mathit{rev}_1$ to $\mathit{rev}_4$ in \autoref{fig:trace}.
$\mathit{rev}_1$ revokes the delegation of \texttt{a} right after it is issued, so it is concurrent with all updates and invalidates roughly $\frac{1}{3}$ of them.
$\mathit{rev}_2$ revokes the delegation of \texttt{a.a} right after it is issued in the middle of the trace, so it is concurrent with half of the updates and invalidates roughly 4\,\% of them.
$\mathit{rev}_3$ is concurrent only with the last update, which it invalidates.
Finally, $\mathit{rev}_4$ is placed after all updates and invalidates none. %
For $\mathit{rev}_1$ to $\mathit{rev}_3$, we compare traces with 10\,000 and 100\,000 updates, whereas $\mathit{rev}_4$ requires no rematerialization and is thus unaffected by trace size.

Our benchmarks do not include actual networks: values are only serialized and deserialized into in-memory data structures.
The per-update bandwidth overhead is modest: each update event gains a 32-byte SHA3-256 commitment and each delta a 32-byte salt, comparable to the 64~bytes per signature and 32~bytes per predecessor pointer of the signed baseline.
Delegation and revocation events have no counterpart without access control, but carry only a few short path strings or a single hash pointer. Assuming both are far rarer than updates, their bandwidth cost is negligible wherever the baseline is viable.

\paragraph{Hardware and Software Setup}
All benchmarks ran on a desktop computer with an AMD Ryzen 9 5950X processor and Arch Linux (kernel 7.2.6), using OpenJDK 27 and the Java Microbenchmark Harness (JMH). Ed25519 signatures were computed with the Bouncy Castle library.

\subsection{Results}
\begin{figure}
  \centering
  \includegraphics{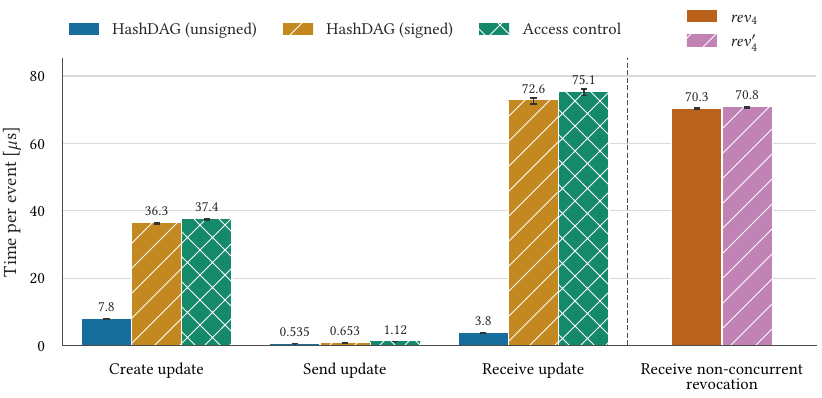}
  \caption{Time to create, send, and receive updates and to receive non-concurrent revocations with a trace of 100\,000 updates. Error bars are JMH's 99.9\,\% confidence intervals over five measured iterations.}
  \label{fig:create-send-receive}
\end{figure}

\paragraph{(1) Creating Updates.}
The first benchmark measures the average time to create and locally apply an update event. \autoref{fig:create-send-receive} shows the average runtimes for a trace of 100\,000 updates in its three left bars. 
The cost is dominated by Ed25519 signing: both signed systems are roughly $30\,\mu\mathrm{s}$ slower than the unsigned baseline, while access control adds only roughly $1\,\mu\mathrm{s}$ on top of the signed baseline.

\paragraph{(2) Sending Updates.}
The second benchmark measures the average time a replica spends sending a single delta to a peer. For the access-control system, this involves verifying that the peer holds a valid capability for reading the delta's path, then serializing the authorization event and the delta payload separately.
The results appear in the second group of \autoref{fig:create-send-receive}. The two baselines behave similarly: since no signatures are computed in this step, the additional metadata carried by signatures accounts for a difference of only about $0.1\,\mu\mathrm{s}$ between them.
The access-control system adds roughly $0.6\,\mu\mathrm{s}$ over both baselines.

\paragraph{(3) Receiving Updates.}
The third benchmark measures the overhead of processing a received update. For the access-control system, this includes deserializing the event and the delta, computing the event hash and verifying the 
commitment and signature, and checking read and write authorization. 
As \autoref{fig:create-send-receive} shows, signature verification dominates: both signed systems perform similarly, with the 
transition from unsigned to signed adding about $70\,\mu\mathrm{s}$, while the access-control system adds only roughly $3\,\mu\mathrm{s}$ on top of the signed baseline.

\begin{figure}
  \centering
  \includegraphics{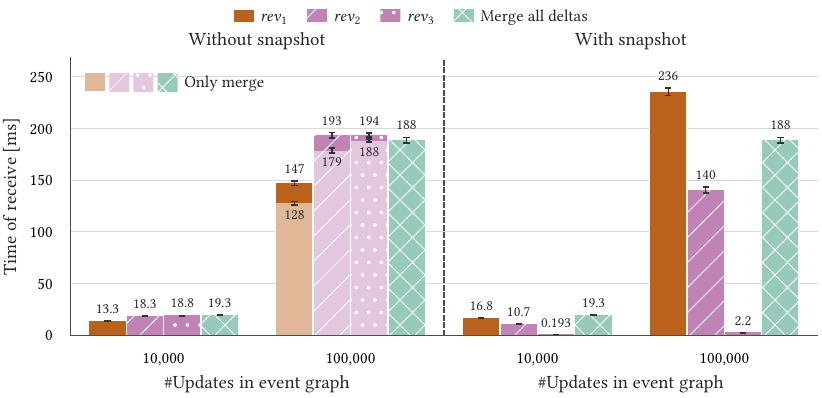}
  \caption{Time for receiving a concurrent revocation event with and without snapshots, for a trace of 10\,000 and 100\,000 events. Error bars are JMH's 99.9\,\% confidence intervals over five measured iterations.}
  \label{fig:receive-revocation-split}
\end{figure}

\paragraph{(4) Applying Revocations.}
Since creating a revocation requires no recomputation, we only measure receiving revocations, for which the baselines have no counterpart.
We first consider the non-concurrent revocation $\mathit{rev}_4$ of \texttt{a} and the analogous revocation $\mathit{rev}_4'$ of \texttt{a.a}, which invalidate no applied updates:
as the rightmost group of \autoref{fig:create-send-receive} shows, receiving them is about as efficient as receiving a regular update.

\autoref{fig:receive-revocation-split} shows the concurrent revocations $\mathit{rev}_1$ to $\mathit{rev}_3$.
Revocations that require rematerialization are significantly more expensive than the other operations (note that the plot is in ms).
Most of this cost is due to merging the remaining, non-revoked deltas (bottom part of each bar), while the top part shows the additional cost of the revocation logic.
As a baseline, \emph{Merge all deltas} shows the time to merge all deltas of the event graph without any revocation, i.e., a full materialization.

The revocations $\mathit{rev}_1$ to $\mathit{rev}_3$ invalidate less and less of the state: $\mathit{rev}_1$ invalidates $\approx\frac{1}{3}$ of the updates, $\mathit{rev}_2$ $\approx4\,\%$, and $\mathit{rev}_3$ a single update.
Consequently, the revocation logic gets cheaper (19\,ms, 14\,ms, and 6\,ms, respectively), while recomputation gets more expensive, as more deltas remain to be merged (128\,ms, 179\,ms, and 188\,ms, respectively, the latter as expensive as merging all deltas).
Overall, every concurrent revocation is significantly more expensive than a non-concurrent one, and costs about as much as a full materialization.

\paragraph{Revocation With Snapshots.}
Since merging dominates the cost of concurrent revocations, snapshots can reduce it.
We use a simple strategy: a single snapshot at the causal stability frontier, i.e., the point seen by all replicas, so that, barring backdating, no revocation can be concurrent with the deltas in the snapshot.
In our benchmarks, this snapshot is taken right before the revocation.

For $\mathit{rev}_1$, the snapshot is effectively empty, so there is no potential for optimization. The approach is even slower (236\,ms vs.\ 147\,ms without snapshots), since deciding which deltas to include is more costly.
$\mathit{rev}_2$ is faster with the snapshot (140\,ms vs.\ 193\,ms), as only roughly half of the deltas need to be merged on top of the snapshot.
For $\mathit{rev}_3$, which invalidates a single update near the end of the trace, the cost falls from 194\,ms to 2\,ms, because partial recomputation amounts to merging a single delta on top of the snapshot.

\paragraph{Discussion of Results.}
The absolute overhead of access control is small in comparison to signature processing.
Creating, sending, and receiving an update together take $\approx 114\,\mu\mathrm{s}$ with access control, only $\approx 4\,\mu\mathrm{s}$ or $\approx 4\,\%$ more than with the signed baseline.

Receiving a revocation that invalidates no applied updates is about as efficient as receiving a regular update.
Revocations that invalidate applied updates, which we assume to be rare and which are not on the critical path, incur two costs.
The access-control logic, i.e., determining the invalidated updates, adds only $6$--$19\,\mathrm{ms}$ at 100\,000 updates.
The dominant cost is rematerialization, i.e., merging the remaining deltas, which depends on the application's \gls{ardt} and grows roughly linearly with the number of updates.
Without backdating, snapshots at the causal stability frontier avoid most of this merging cost.

We conclude from these observations that GUARDTs introduce acceptable overheads in open networks with latencies above $1\,\mathrm{ms}$ and human-scale operation frequencies.

\section{Related Work}\label{chap:related work}

\citet{renauxSecureRDTsEnforcing2023}
propose an access-control scheme for synchronizing 
JSON CRDTs between untrusted clients and a trusted central server that enforces access control. 
Permissions are specified using a policy specification language based on the paths in the JSON CRDTs.
Their use of paths is similar to that of GUARDTs, but their policy language is more expressive.
Their approach, however, requires a single trusted centralized server, is tied to JSON CRDTs, and assumes a static policy.

\citet{raultDistributedAccessControl2022} %
propose access control for CRDTs using a policy CRDT in conjunction with a separate data-storing CRDT.
The merge function for the policy CRDT resolves conflicts using a combination of 
the authors' role hierarchy and a deterministic tie-breaker.
To avoid divergence of state caused by concurrent policy and data operations,
their model requires undo and redo operations for both CRDTs,
which are used to roll back conflicting operations.
In contrast to our approach, which supports fine-grained access control for arbitrary 
ARDT compositions, their scheme is limited to specific CRDT constructions
and is not Byzantine fault-tolerant.

Weber et al.~\cite{weberAccessControlWeakly2016, weberACGreGateFrameworkPractical2018}
propose access control for a weakly consistent information system in which mutually trusted servers replicate a dataset and enforce access control towards external clients. The servers also replicate a policy CRDT that resolves concurrent policy versions by their intersection, granting only permissions present in all of them.
While their setting differs from ours -- the data-management layer versus the application-programming layer --
the two approaches share correctness criteria:
policy updates and dependent operations must be handled in causal order (\emph{protection relation}),
and conflicting policy updates must be resolved in favor of restrictions (\emph{conservative conflict resolution}).
For reads, however, conservative resolution of concurrent revocations is impossible without a total order~\cite{yanakievaImpossibilityConfidentialityIntegrity2021}, so we enforce only causally preceding ones.

\citet{wobberPolicybasedAccessControl2010} were the first to propose eventually consistent access control for mutually distrustful replicas without central synchronization infrastructure.
Their approach authenticates updates with signatures and detects unauthorized writes, but neither tolerates Byzantine replicas nor ensures convergence itself.
Policies are specified in a separate logical assertion language over a flat namespace of item labels, independent of the structure of the data.

\citet{jacobBestKnowledgeBelief2025} present a theoretical model for eventually consistent access control (ECAC) that draws inspiration from the Matrix system. The model derives a mutable total order of events over which authorization policies can be specified.
GUARDTs do not rely on such a derived total order but instead exploit the immutable causal order of the event graph.
Crucially, the model by Jacob and Hartenstein targets the replication layer, not the application layer. As such, it leaves open how policies are structured relative to application data, while we derive fine-grained, path-based permissions from the application data structure.

\citet{yanakievaImpossibilityConfidentialityIntegrity2021} study access control for partition-tolerant distributed file systems with POSIX-like semantics. They show that, without a total order, a file system cannot enforce standard POSIX access control, supporting our conclusion that revocations require more nuanced semantics for concurrent reads and writes.

\section{Conclusion}\label{chap:conclusion}

We presented GUARDTs, replicated data types with decentralized access control for strongly eventually consistent systems with mutually distrustful replicas.
Unlike existing approaches, which require a trusted coordinator or static policies, assume cooperative replicas, or are tied to specific data-type constructions, GUARDTs achieve all four properties that principled decentralized access control for RDTs demands.

Path-based capabilities derived from the tree structure of ARDTs express fine-grained read and write permissions over arbitrary application data types (generality).
Replicas can delegate and transitively revoke capabilities at runtime (policy evolution). %
Authorization events are recorded in a signed HashDAG replicated as a CRDT among all replicas, providing tamper-evident, causally ordered evidence for local authorization decisions that remain correct under Byzantine behavior (Byzantine resilience).
Every replica enforces access control by filtering outgoing deltas and verifying authority on incoming ones, with no central policy server (decentralization).

We formally proved that GUARDTs guarantee confidentiality, integrity, and availability despite concurrent policy changes and Byzantine replicas. Empirically, the total per-operation overhead of access control over a signed HashDAG baseline is roughly 4\,\%, confirming that the mechanism is practical.
The dominant cost of revocations that invalidate already-applied updates is rematerialization of the affected state. Snapshots at the causal stability frontier can reduce this overhead.

As interesting avenues for future work, we see using encryption as a complementary or alternative means of enforcing read access.
Another extension is to automatically derive the set of paths that an application-level operation requires, such as appending an element to a list, which modifies multiple paths,
or to use more expressive path descriptors inside the capabilities.

\bibliographystyle{ACM-Reference-Format}
\bibliography{bibliography}

\end{document}